\documentclass[prl,twocolumn,aps,psfig,showpacs,nofootinbib,nobibnotes,superscriptaddress,preprintnumbers]{revtex4}

\usepackage{epsfig}
\usepackage{graphics}
\usepackage{bm}
\usepackage{color}

\newcommand{\uu}{\bm{u}}
\newcommand{\vv}{\bm{v}}
\newcommand{\BB}{\bm{B}}
\newcommand{\AAA}{\bm{A}}
\newcommand{\JJ}{\bm{J}}

\newcommand\kk{\mathbf{k}}

\newcommand{\ff}{\bm{f}}
\newcommand{\xx}{\bm{x}}
\newcommand{\nab}{\mbox{\boldmath $\nabla$} {}}

\def\urms{u_{\rm rms}}
\def\cs{c_{\rm s}}
\def\tauA{\tau_{\rm A}}
\def\vA{v_{\rm A}}
\def\vAz{v_{\rm A0}}
\def\kM{k_{\rm M}}

\newcommand{\bra}[1]{\langle #1\rangle}
\newcommand{\Fig}[1]{Fig.\ \ref{#1}}

\newcommand{\Sec}[1]{Sec.\ \ref{#1}}
\def\EK{E_{\rm K}}
\def\EM{E_{\rm M}}

\def\EWT{E_{\rm WT}}

\def\EEK{{\cal E}_{\rm K}}
\def\EEM{{\cal E}_{\rm M}}
\def\HHM{{\cal H}_{\rm M}}
\def\Pm{\mbox{\rm Pr}_{\rm M}}

\def\Lu{\mbox{\rm Lu}}
\def\Rey{\mbox{\rm Re}}
\def\epsK{\epsilon_{\rm K}}
\def\epsM{\epsilon_{\rm M}}
\def\xiM{\xi_{\rm M}}
\def\xiTay{\xi_{\rm Tay}}

\newcommand{\yprd}[3]{, Phys.\ Rev.\ D {\bf #2}, #3 (#1).}
\newcommand{\ypre}[3]{, Phys.\ Rev.\ E {\bf #2}, #3 (#1).}
\newcommand{\yprl}[3]{, Phys.\ Rev.\ Lett.\ {\bf #2}, #3 (#1).}
\newcommand{\yjfm}[3]{, J.\ Fluid Mech.\ {\bf #2}, #3 (#1).}

\def\half{{1\over2}}

\newcommand{\EQA}{\begin{eqnarray}}
\newcommand{\ENA}{\end{eqnarray}}
\newcommand{\EQ}{\begin{equation}}
\newcommand{\EN}{\end{equation}}
\newcommand{\ii}{{\rm i}}
\newcommand{\const}{{\rm const}  {}}
\newcommand{\rr}{\mbox{\boldmath $r$} {}}
\newcommand{\Tab}[1]{Table~\ref{#1}}
\newcommand{\Eq}[1]{Equation~(\ref{#1})}
\newcommand{\Eqs}[2]{Equations~(\ref{#1}) and~(\ref{#2})}
\newcommand{\Eqss}[2]{Equations~(\ref{#1})--(\ref{#2})}
\newcommand{\zzz}{\hat{\mbox{\boldmath $z$}} {}}
\newcommand{\oo}{\mbox{\boldmath $\omega$} {}}
\newcommand{\bb}{\bm{b}}
\newcommand{\jj}{\mbox{\boldmath $j$} {}}
\newcommand{\aaaa}{\mbox{\boldmath $a$} {}}
\newcommand{\eee}{\hat{\mbox{\boldmath $e$}} {}}
\newcommand{\ssigma}{\mbox{\boldmath $\sigma$} {}}
\newcommand{\hatAA}{\hat{\bm{A}}}
\def\xiK{\xi_{\rm K}}
\def\kK{k_{\it K}}

\newcommand{\ypf}[3]{, Phys. Fluids {\bf #2}, #3 (#1).}
\newcommand{\yapj}[3]{, Astrophys. J. {\bf #2}, #3 (#1).}
\newcommand{\yan}[3]{, Astron. Nachr. {\bf #2}, #3 (#1).}
\newcommand{\ycsf}[3]{, Chaos, Solitons \& Fractals {\bf #2}, #3 (#1).}
\newcommand{\yjour}[4]{, #2 {\bf #3}, #4 (#1).}

\begin{document}

\title{Nonhelical inverse transfer of a decaying turbulent magnetic field}

\date{\today,~ $ $Revision: 1.4 $ $}
\preprint{NORDITA-2014-42}

\author{Axel Brandenburg}
\email{brandenb@nordita.org}
\affiliation{Nordita, KTH Royal Institute of Technology and Stockholm University,
Roslagstullsbacken 23, 10691 Stockholm, Sweden}
\affiliation{Department of Astronomy, AlbaNova University Center,
Stockholm University, 10691 Stockholm, Sweden}

\author{Tina Kahniashvili}
\email{tinatin@phys.ksu.edu}
\affiliation{The McWilliams Center for
Cosmology and Department of Physics, Carnegie Mellon University,
5000 Forbes Ave, Pittsburgh, PA 15213, USA}
\affiliation{Department of Physics, Laurentian University, Ramsey
Lake Road, Sudbury, ON P3E 2C, Canada} \affiliation{Abastumani Astrophysical Observatory, Ilia State
University, 3-5 Cholokashvili Ave, Tbilisi, GE-0194, Georgia}

\author{Alexander G.\ Tevzadze}
\email{aleko@tevza.org}
\affiliation{Faculty of Exact and Natural
Sciences, Tbilisi State University, 1 Chavchavadze Ave., Tbilisi,
0128, Georgia}

\begin{abstract}

In the presence of magnetic helicity, inverse transfer from small to
large scales is well known in magnetohydrodynamic (MHD) turbulence and has
applications in astrophysics, cosmology, and fusion plasmas.
Using high resolution direct numerical simulations of magnetically
dominated self-similarly decaying MHD turbulence, we report
a similar inverse transfer even in the absence of magnetic helicity.
We compute for the first time spectral energy transfer rates to show
that this inverse transfer is about half as strong as with helicity, but
in both cases the magnetic gain at large scales results from velocity
at similar scales interacting with smaller-scale magnetic fields.
This suggests that both inverse transfers are a consequence of
a universal mechanisms for magnetically dominated turbulence.
Possible explanations include inverse cascading of the mean squared vector
potential associated with local near two-dimensionality and the
shallower $k^2$ subinertial range spectrum of kinetic energy
forcing the magnetic field with a $k^4$ subinertial range
to attain larger-scale coherence.
The inertial range shows a clear $k^{-2}$ spectrum and is
the first example of fully isotropic magnetically dominated
MHD turbulence exhibiting weak turbulence scaling.
\end{abstract}


\maketitle

The nature of magnetohydrodynamic (MHD) turbulence has received
significant attention in recent years \cite{Ber14}.
Whenever plasma is ionized, it is electrically conducting and
Kolmogorov's turbulence theory \cite{kol} has to be replaced
by an appropriate theory for MHD turbulence \cite{b1}.
This becomes relevant under virtually all astrophysical circumstances.
However, the universal character of MHD turbulence is debated and several
fundamental questions remain unanswered: how do kinetic and magnetic energy
spectra look like and are they similar?
How does this depend on the magnetic Prandtl number, $\Pm=\nu/\eta$,
i.e., the ratio of kinematic viscosity and magnetic diffusivity?
What is the role of the Alfv\'en effect, i.e., how does the presence
of a finite Alfv\'en speed $\vA$ enter the expression for the
turbulent energy spectrum?

If the spectral properties of MHD turbulence are governed solely
by the rate of energy transfer $\epsilon$, we know from
dimensional arguments that the spectrum must scale as
$E(k)\sim\epsilon^{2/3}k^{-5/3}$ with wavenumber $k$.
However, MHD turbulence becomes increasingly anisotropic toward small scales
\cite{GS95}, so the spectrum $E(k_\perp,k_\|)$ depends on the wavenumbers
perpendicular and parallel to the magnetic field $\BB$, and is essentially
given by $\epsilon^{2/3}k_\perp^{-5/3}$, so most of the energy cascades
perpendicular to $\BB$.

In the case of forced turbulence,
direct numerical simulations (DNS) show similar spectra
both with imposed \cite{Ber14} and dynamo-generated \cite{HBD04} fields.
However, when $\BB$ is decaying, the result 
depends on the value of the initial ratio $\vA/\urms$
of root mean square (rms) Alfv\'en speed to rms turbulent velocity.
Recent DNS \cite{Lee2010} found numerical evidence for three different
scalings: Iroshnikov--Kraichnan scaling \cite{Iro+Kra} proportional to
$(\epsilon\vA)^{1/2}k^{-3/2}$ for $\vA/\urms=0.9$,
Goldreich--Sridhar scaling \cite{GS95} proportional to
$\epsilon^{2/3}k_\perp^{-5/3}$ for $\vA/\urms=1.3$, and
weak turbulence scaling \cite{Gal00} proportional to
$(\epsilon\vA k_\|)^{1/2}k_\perp^{-2}$ for $\vA/\urms=2.0$;
see Ref.~\cite{BN11} for a comparison of these three scalings.
However, their physical interpretation is subject
to criticism in that dynamic alignment between $\uu$ and
$\BB$ can be responsible for the shallower $k^{-3/2}$ scaling
\cite{MCB06} and the $k^{-2}$ scaling could also be caused by a
dominance of discontinuities \cite{DA14}.

It is usually taken for granted that
for non-helical turbulence, energy is cascading toward small scales.
An inverse cascade has so far only been found for helical turbulence
\cite{b1,PFL} and was confirmed in DNS \cite{BP99,a1,B01}.
It is evident that this requires significant scale separation,
$k_0/k_1\gg1$, where $k_0$ is the wavenumber of the peak of the spectrum
and $k_1=2\pi/L$ is the minimal wavenumber of the domain of size $L$.
Since an inverse transfer was not expected to occur in the absence of
helicity, most previous work did not allow for $k_0/k_1\gg1$.
However, when $k_0/k_1$ is moderate, some inverse cascading
was found \cite{a1}.
The present work shows that this behavior is genuine and more
pronounced at higher resolution, larger Reynolds numbers and larger $k_0/k_1$.

\begin{figure}[t!]
\includegraphics[width=.9\columnwidth]{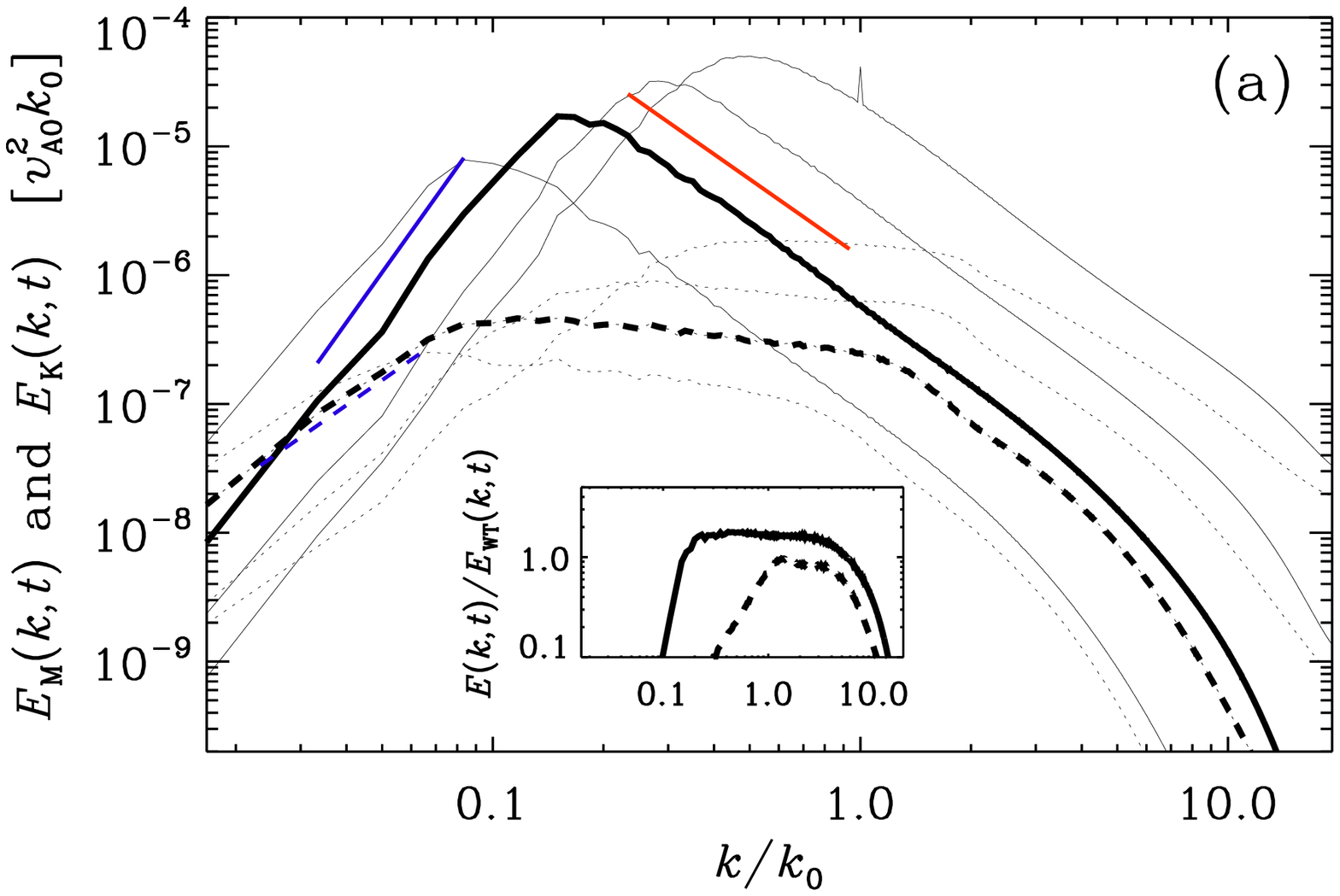}
\includegraphics[width=.9\columnwidth]{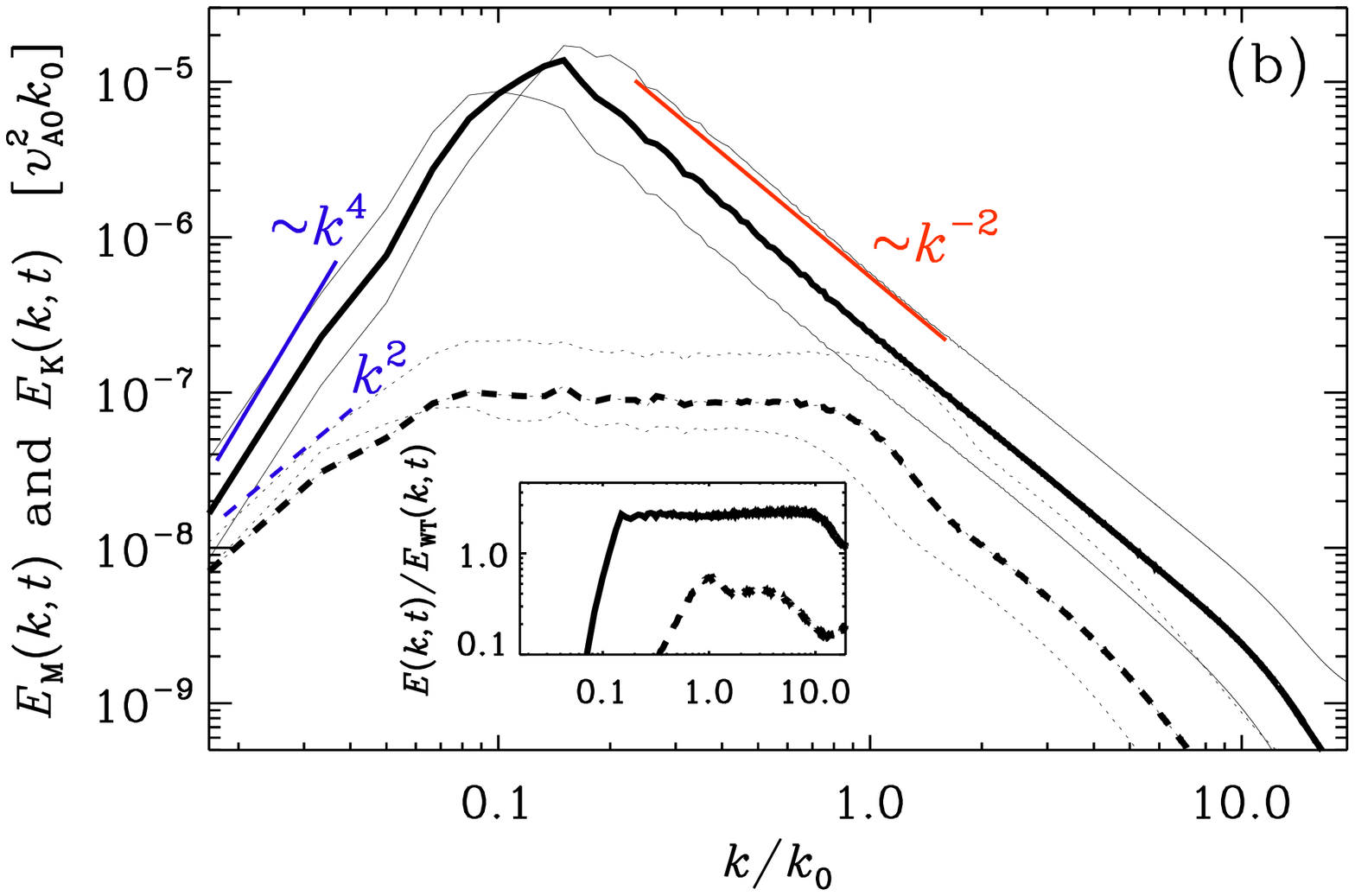}
\caption[]{(Color online)
(a) Magnetic (solid lines) and kinetic (dashed lines) energy
spectra for Run~A at times $t/\tauA=18$, 130, 450, and 1800;
the time $t/\tauA=450$ is shown as bold lines.
The straight lines indicate the slopes $k^4$ (solid, blue),
$k^2$ (dashed, blue), and $k^{-2}$ (red, solid).
(b) Same for Run~B, at $t/\tauA=540$, 1300, and 1800, with $t/\tauA=1300$
shown as bold lines.
The insets show $\EM$ and $\EK$ compensated by $\EWT$.
}\label{non-helical-spectrum}
\end{figure}

We solve the compressible MHD equations for $\uu$,
the gas density $\rho$ at constant sound speed $\cs$,
and the magnetic vector potential $\AAA$, so $\BB=\nab\times\AAA$.
Following our earlier work \cite{kbtr10,Tevzadze:2012kk,Kahniashvili:2012uj},
we initialize our decaying DNS by restarting them from
a snapshot of a driven DNS, where a random forcing
was applied in the evolution equation for $\AAA$ rather than $\uu$.
To allow for sufficient scale separation, we take $k_0/k_1=60$.
We use the {\sc Pencil Code} \cite{B-pencil} at a resolution of
$2304^3$ meshpoints on 9216 processors.
The code uses sixth order finite differences and a third order
accurate time stepping scheme.

\begin{figure}[t!]
\includegraphics[width=\columnwidth]{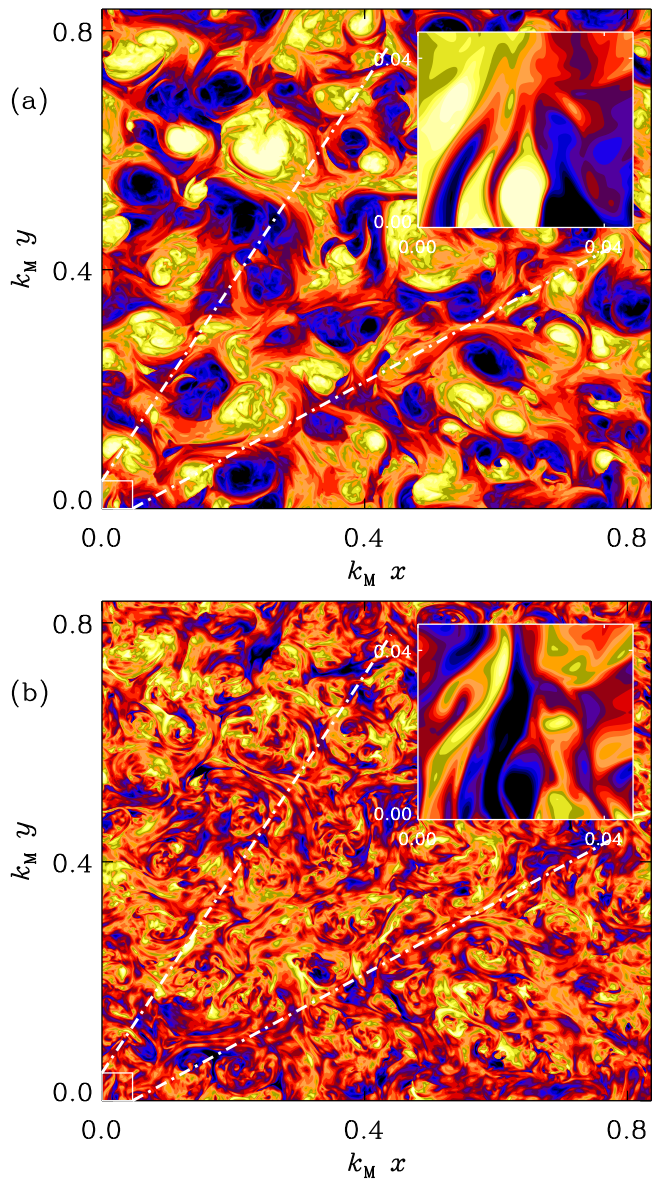}
\caption[]{(Color online)
Contours of (a) $B_z(x,y)$ and (b) $u_z(x,y)$ for Run~A.
The insets show a zoom into the small square in the lower left corner.
}\label{pslice}
\end{figure}

\begin{figure*}[t!]
\includegraphics[width=.95\textwidth]{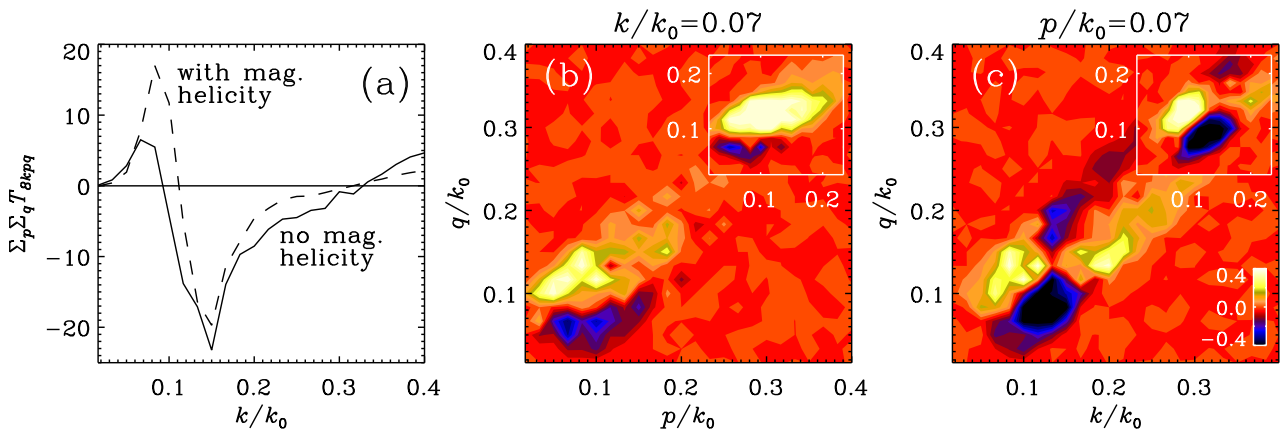}
\caption[]{(Color online)
Spectral transfer function $T_{kpq}$,
(a) as a function of $k$ and summed over all $p$ and $q$,
(b) as a function of $p$ and $q$ for $k/k_1=4$, and
(c) as a function of $k$ and $q$ for $p/k_1=4$.
The dashed line in (a) and the insets in (b) and (c)
show the corresponding case for a DNS with helicity; both for $\Pm=1$.
}\label{pTM_kpq_comp}
\end{figure*}

Our magnetic and kinetic energy spectra are normalized such that
$\int \EM(k,t)\,dk=\EEM (t)=\vA^2/2$ and
$\int \EK(k,t)\,dk=\EEK(t)=\urms^2/2$ are magnetic and kinetic
energies per unit mass.
The magnetic integral scale is defined as
$\xiM=\kM^{-1}(t)=\int k^{-1} \EM(k,t)\,dk/\EEM(t)$.
Time is given in initial Alfv\'en times $\tauA=(\vAz k_0)^{-1}$,
where $\vAz=\vA(0)$.
In \Fig{non-helical-spectrum} we show $\EM(k,t)$ and $\EK(k,t)$
for Runs~A and B (restarted from A at $t/\tauA=450$)
with $\Pm=1$ and $10$, respectively,
and in \Fig{pslice} slices $B_z(x,y)$ and $u_z(x,y)$ at $z=0$ at the
last time Run~A.
We find an inertial range with weak turbulence scaling,
\begin{equation}
\EWT(k,t)=C_{\rm WT}(\epsilon\vA \kM)^{1/2}k^{-2},
\end{equation}
where $\kM^{-1}(t)=\int k^{-1} \EM(k,t)\,dk/\EEM(t)$
is the integral scale and $\kM$ has been used in place of $k_\|$.
The prefactor is $C_{\rm WT}\approx1.9$ for $\Pm=1$ and
$\approx2.4$ for $\Pm=10$; see the insets.
In agreement with earlier work \cite{b1,Tevzadze:2012kk},
$\EEM$ decays like $t^{-1}$.

At small wavenumbers the $k^4$ and $k^2$ subinertial ranges
respectively for $\EM(k,t)$ and $\EK(k,t)$ are carried over
from the initial conditions.
The $k^4$ Batchelor spectrum is in agreement with the causality
requirement \cite{cd01,dav} for the divergence-free vector field $\BB$.
The velocity is driven entirely by the magnetic field and follows
a white noise spectrum, $\EK (k) \propto k^2$ \cite{dav}.
The resulting difference in the scaling implies that,
although magnetic energy dominates over kinetic, the two spectra
must cross at sufficiently small wavenumbers.
This idea may also apply to incompressible \cite{BL14}
and relativistic \cite{Zra14} simulations, where
inverse nonhelical transfer has recently been confirmed.

To quantify the nature of inverse transfer
we show in \Fig{pTM_kpq_comp} representations of
the spectral transfer function $T_{kpq}=\bra{\JJ^k\cdot(\uu^p\times\BB^q)}$
and compare with the corresponding helical case of
Ref.~\cite{Kahniashvili:2012uj}, but with $1024^3$ mesh points
and at a comparable time.
Here, the superscripts indicate the radius of a shell in wavenumber space
of Fourier filtered vector fields; see Ref.~\cite{B01} for such an analysis
in driven helical turbulence.
The transfer function $T_{kpq}$ quantifies the gain of magnetic energy
at wavenumber $k$ from interactions of velocities at wavenumber $p$ and
magnetic fields at wavenumber $q$.
\Fig{pTM_kpq_comp}(a) shows a gain for $k/k_0<0.1$, which is about half of
that for the helical case.
The corresponding losses for $k/k_0>0.1$ are about equal in the two cases.
In both cases, the magnetic gain at $k/k_0=0.07=4/60$ results from
$\uu^p$ with $0<p/k_0<0.2$ interacting with $\BB^q$ at $q/k_0>0.1$;
see the light yellow shades in \Fig{pTM_kpq_comp}(b).
Note that work done by the Lorentz force is
$\bra{\uu^p\cdot(\JJ^k\times\BB^q)}=-T_{kpq}$.
Thus, negative values of $T_{kpq}$ quantify the gain of {\it kinetic} energy at
wavenumber $p$ from interactions of magnetic fields at wavenumbers $k$ and $q$.
Blue dark shades in \Fig{pTM_kpq_comp}(c) indicate therefore that the gain of
kinetic energy at $p/k_0=0.07$ results from magnetic interactions at
wavenumbers $k$ and $q$ of around $0.1\,k_0$.
These results support the interpretation that the increase of spectral power
at large scales is similar to the inverse transfer in the helical case;
see \cite{suppl} for information concerning the total energy transfer.

\begin{figure}[t!]
\includegraphics[width=\columnwidth]{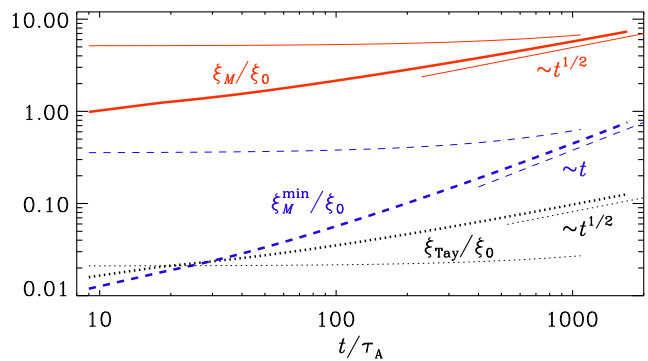}
\caption[]{
Time evolution of $\xiM=\kM^{-1}$ and $\xiM^{\min}$,
as well as the Taylor microscale $\xiTay$.
Fat (thin) lines are for Run~A (B).
} \label{non-helical-decay}
\end{figure}

To exclude that the inverse energy transfer is a consequence of the
invariance of magnetic helicity, $\HHM(t)=\langle\AAA\cdot\BB\rangle$,
we compare $\xiM$ with its lower bound $\xiM^{\min}=|\HHM|/2\EEM$
\cite{Tevzadze:2012kk}; see Fig.~\ref{non-helical-decay}.
In nonhelical MHD turbulence, $\xiM$ is known to grow like $t^{1/2}$
\cite{b1,Tevzadze:2012kk}.
Even though the initial condition was produced with nonhelical plane
waves, we find $\HHM\neq0$ due to fluctuations.
Since $\HHM$ is conserved and $\EEM$ decays like $t^{-1}$
\cite{b1,Tevzadze:2012kk}, $\xiM^{\min}$
grows linearly and faster than $\xiM\sim t^{1/2}$, so they will meet at
$t/\tauA=10^5$ and then continue to grow as $t^{-2/3}$
\cite{b1,Tevzadze:2012kk}, but at $t/\tauA=10^3$ this cannot explain
the inverse transfer.
By contrast, we cannot exclude the possibility of the quasi
two-dimensional mean squared vector potential, $\bra{\AAA_{\rm 2D}^2}$,
being approximately conserved \cite{suppl}.
This could explain the $\xiM\sim t^{1/2}$ scaling and the inverse transfer
if the flow was locally two-dimensional \cite{Pou78}.

Since $\urms$, $\vA$, and $\kM$ are all proportional to $t^{-1/2}$
the decay is self-similar in such a way that the Reynolds and Lundquist
numbers, $\Rey=\urms/\nu\kM$ and $\Lu=\vA/\eta\kM$, remain constant.
Since $\EEK\ll\EEM$, the dissipated energy comes predominantly
from $-d\EEM/dt$, and yet a substantial fraction of it is used
to drive kinetic energy by performing work on the Lorentz force.
Nevertheless, the viscous to magnetic dissipation ratio
$\epsK/\epsM$ increases only by a factor of 1.8 as $\Pm$ increases
from 1 to 10; see Table~\ref{Tab}.
This is less than for kinetically driven MHD turbulence,
where $\epsK/\epsM\propto\Pm^{n}$ with $n=0.3$--$0.7$ \cite{B14}.
Therefore, $\epsM$ is here larger than in driven MHD turbulence,
where large $\Lu$ can still be tolerated.
This suggests that Run~B may be under-resolved, which might also explain
why it did not reach asymptotic scaling in \Fig{non-helical-decay}.

\begin{table}[t]\caption{
Comparison of relative dissipation rates, energies
and other parameters for the two simulations discussed.
}\vspace{12pt}\centerline{\begin{tabular}{ccccccccc}
Run & $\Pm$ & $\vAz/\cs$ & $\urms/\vA$ & $\Lu$ & $\Rey$ & 
$\epsK/\epsM$ & $\epsM/\epsilon$ & ~$\epsK/\epsilon$ \\
\hline
A & ~1 & 0.15 & 0.36 & ~700 & 230 & 0.52 & 0.66 & 0.34 \\
B & 10 & 0.03 & 0.21 & 6300 & 130 & 0.93 & 0.52 &~0.48 
\label{Tab}\end{tabular}}\end{table}

In summary, we have shown that inverse transfer is a ubiquitous
phenomenon of both helical and non-helical MHD.
For helical MHD, this has been well known for nearly four
decades \cite{PFL}, but for nonhelical MHD there have only
been some low resolution DNS \cite{a1,Kahniashvili:2012uj}.
Our DNS confirm an early finding by Olesen \cite{Ole97} that this inverse
transfer occurs for all initial spectra that are sufficiently steep.
His argument applies to hydrodynamic and MHD turbulence if the two spectra
are parallel to each other.
In our case, however, owing to the shallower $k^2$ spectrum of kinetic energy,
kinetic energy always dominates over magnetic at large enough length scales.
Either this or the near-conservation of $\bra{\AAA_{\rm 2D}^2}$ could
be responsible for inverse transfer in magnetically dominated turbulence.
This process is significant for cosmology and astrophysics \cite{Zra14}, with
applications not only to primordial magnetic fields, but also to ejecta
from young stars, supernovae, and active galactic nuclei \cite{BDLK13}.

Our results support the idea of the weak turbulence $k^{-2}$ scaling
for strong magnetic field that is here for the first time globally
isotropic and not an imposed one \cite{PB08}.
At small scales, however, approximate equipartition is still possible.
The decay is slower than for usual MHD turbulence
which is arguably governed by the Loitsyansky invariant \cite{Dav00}.
Future investigations of the differences between these types
of turbulence are warranted \cite{suppl}.
Interestingly, the extended plateau in the velocity spectrum around
the position of the magnetic peak may be important for producing
observationally detectable broad gravitational wave spectra \cite{KCGMR08}.

\acknowledgements

We appreciate useful discussions with A.\ Neronov.
Computing resources have been
provided by the Swedish National Allocations Committee at the
Center for Parallel Computers at the Royal Institute of Technology
and by the Carnegie Mellon University Supercomputer Center.
We acknowledge support from
the Swedish Research Council grants 621-2011-5076 and 2012-5797,
the European Research Council AstroDyn Project 227952,
the Research Council of Norway FRINATEK grant 231444,
the Swiss NSF grant SCOPES IZ7370-152581,
the NSF grant AST-1109180, and
the NASA Astrophysics Theory Program grant NNXlOAC85G.
A.B.\ and A.T.\ acknowledge the hospitality of the McWilliams Center
for Cosmology.

\newcommand{\yjourS}[4]{, #2 {\bf #3}, #4 (#1)}
\newcommand{\ysov}[3]{, Sov. Astron. {\bf #2}, #3 (#1).}
\newcommand{\ypp}[3]{, Phys. Plasmas {\bf #2}, #3 (#1).}
\newcommand{\yanSS}[3]{Astron. Nachr. {\bf #2}, #3 (#1).}
\newcommand{\ysovS}[3]{, Sov. Astron. {\bf #2}, #3 (#1)}
\newcommand{\ypfS}[3]{, Phys. Fluids {\bf #2}, #3 (#1)}
\newcommand{\yapjl}[3]{, Astrophys. J. Lett. {\bf #2}, #3 (#1).}
\newcommand{\yapjS}[3]{, Astrophys. J. {\bf #2}, #3 (#1)}
\newcommand{\yapjSS}[3]{Astrophys. J. {\bf #2}, #3 (#1)}
\newcommand{\yjgr}[3]{, J. Geophys. Res. {\bf #2}, #3 (#1).}
\newcommand{\yjgrS}[3]{, J. Geophys. Res. {\bf #2}, #3 (#1)}
\newcommand{\ypreS}[3]{, Phys.\ Rev.\ E {\bf #2}, #3 (#1)}
\newcommand{\yprlS}[3]{, Phys.\ Rev.\ Lett.\ {\bf #2}, #3 (#1)}
\newcommand{\ymn}[3]{, Mon.\ Not.\ R.\ Astron.\ Soc.\ {\bf #2}, #3 (#1).}


\appendix
\newpage
\LARGE
\noindent{Supplemental Material}
\normalsize
\vspace{4mm}

\noindent
The subject of magnetically dominated decaying MHD turbulence is relevant
to the early Universe and has received enhanced attention in recent years.
The focus of our Letter \cite{SM_BKT14} is the investigation of inverse transfer
of magnetic energy at the expense of kinetic energy at intermediate scales.
In this Supplemental Material, we present additional details
of the resulting turbulence regarding initial conditions, decay rate,
the mechanisms for inverse transfer, including the question
of local two-dimensionality of the turbulence,
and the spectral energy transfer functions.

The simulations discussed in the Letter are motivated by applications
to the early Universe such as the time of the electroweak phase transition.
As shown in Ref.~\cite{SM_BEO96}, the usual MHD equations can be applied
when time $t$ being replaced by the conformal time
$\tilde{t}=\int d t/R(t)$, where $R(t)$ is the scale factor
in the assumed flat, isotropic, homogeneous Universe
as described by the Robertson-Walker metric.

\section{Initial condition}

Initial conditions can be obtained either as a result of an earlier
turbulence simulation driven by monochromatic driving or the fields
can be synthesized with given power spectra and random phases.
In the following, we describe and compare these different cases.

\subsection{Via monochromatic driving}

\begin{figure}[h!]
\includegraphics[width=\columnwidth]{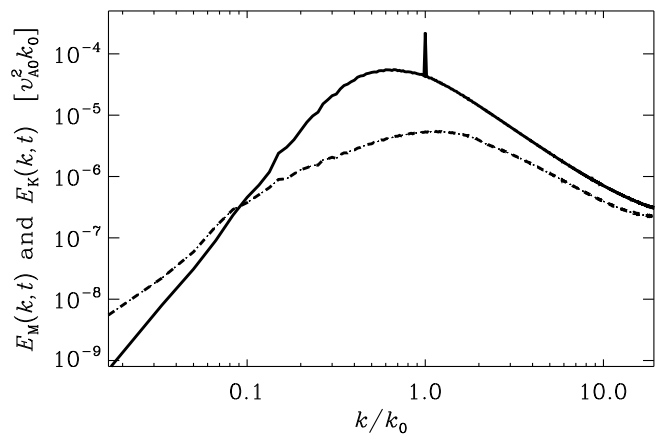}
\caption[]{
Magnetic (solid lines) and kinetic (dashed lines) energy
spectra for the initial condition of Run~A.
}\label{pkt2304_hel_short2304pm1_kf60b_initial}
\end{figure}

Our goal is to have an initial condition that quickly leads to
self-similar decay.
Earlier experience \cite{SM_kbtr10,SM_Tevzadze:2012kk,SM_Kahniashvili:2012uj}
has shown that this is easily achieved by using
a snapshot from a turbulence simulation that was driven with stochastic
monochromatic forcing in the equation for the magnetic vector potential.
The resulting initial condition used in our present work is shown in
\Fig{pkt2304_hel_short2304pm1_kf60b_initial}.
It shows approximate $k^2$ and $k^4$ subinertial ranges for kinetic
and magnetic energy spectra, respectively.
Both spectra are maintained also at later times in such a way that
they gradually shift upward with time (see Fig.~1 of the Letter).

\begin{figure}[h!]\begin{center}
\includegraphics[width=\columnwidth]{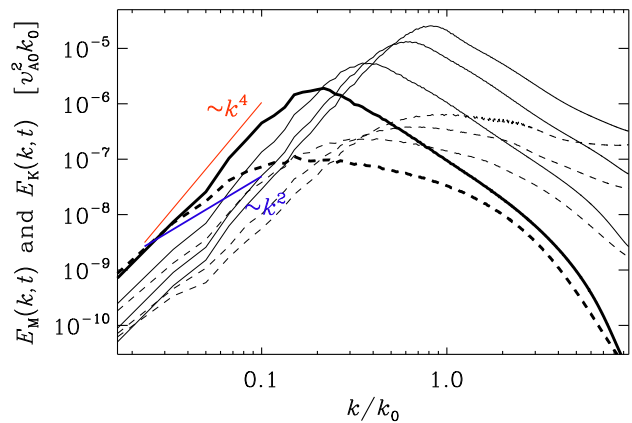}
\end{center}\caption[]{
Like \Fig{pkt2304_hel_short2304pm1_kf60b_initial}, but with an
initial $k^4$ spectrum for the magnetic energy using random phases.
}\label{pkt1152_MKol1152a5}\end{figure}

In the present simulations (Run~A of the Letter),
both magnetic and kinetic energies show a slight uprise of power
near the Nyquist wavenumber, $k_{\rm Ny}=\pi/\delta x$, where $\delta x$
is the mesh spacing.
This indicates that the resolution is only marginal for the
Reynolds number chosen here.
However, during the subsequent decay calculation, after several
Alfv\'en times, this excess power at $k_{\rm Ny}$ disappears,
as is seen in Fig.~1 of the Letter.

\subsection{Via random phases}

An alternative mechanism of producing initial conditions is
to generate a vector field in wavenumber space with a given
spectrum and random phases.
In \Fig{pkt1152_MKol1152a5} we show an example where we have
for magnetic energy a $k^4$ spectrum for $k<k_0$ and $k^{-5/3}$
for $k>k_0$, but zero kinetic energy.
Our initial velocity is zero and the initial vector potential in Fourier space
is $\hat{A}_{j}(\kk)$ such that for all three components $j$ are given by
\EQ
k\hat{A}_{j}(\kk)=A_0{(k/k_0)^{n_1/4-1/2}\over
[1+(k/k_0)^{n_1-n_2}]^{1/4}}\,e^{\ii\phi(\kk)},
\EN
where $n_1=4$ and $n_2=-5/3$ are the exponents of the related
magnetic energy spectrum, $\phi(\kk)$ are random phases,
$A_0$ is the amplitude, and $k=|\kk|$.
We choose $k_0/k_1=60$ and run with $\nu=5\times10^{-6}$
at $1152^3$ meshpoints, which is slightly more dissipative
than the runs reported in the Letter with $\nu=2\times10^{-6}$
using $2304^3$ meshpoints.
In \Fig{pkt1152_MKol1152a5} we show the times $t/\tauA=10$, 50, 200,
and 900, where $\tauA=(\vAz k_0)^{-1}$ is the initial Alfv\'en time.

At very early times ($t\approx\tauA$), a $k^4$ kinetic energy spectrum
develops, which is consistent with the causality constraint, but after
several hundred Alfv\'en times the spectrum becomes gradually shallower
and approaches a $k^2$ subinertial range.
However, unlike the initial condition shown in
\Fig{pkt2304_hel_short2304pm1_kf60b_initial}, the magnetic field
is continuously decaying and the integral scale is increasing,
which is the reason why the $k^2$ subinertial range is less
strongly developed in \Fig{pkt1152_MKol1152a5}. Nevertheless, the
magnetic spectrum shows again clear inverse transfer, although it is
initially somewhat slower, as can be expected given the time it takes
to build up the $k^2$ velocity spectrum.

\subsection{Steeper initial spectra}

\begin{figure}[h!]\begin{center}
\includegraphics[width=\columnwidth]{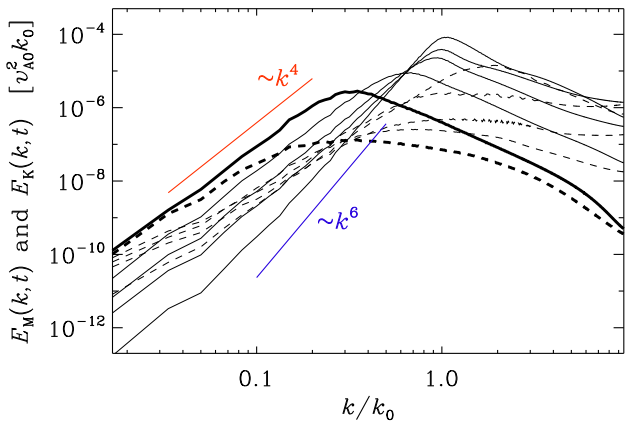}
\end{center}\caption[]{
Like \Fig{pkt1152_MKol1152a5}, but with an initial $k^6$ spectrum
for the magnetic energy using random phases.
}\label{pkt1152_MKol1152_k6b}\end{figure}

If we start with a magnetic energy spectrum steeper than $k^4$,
the spectrum quickly changes into a $k^4$.
This is demonstrated in \Fig{pkt1152_MKol1152_k6b}, where we start
with an initial $k^6$ spectrum, followed by a $k^{-5/3}$ subrange.
We show the times $t/\tauA=1$, 5, 20, 80, and 400, and see that already
at $t/\tauA=20$ the subinertial range has nearly a $k^4$ subrange.

\section{Evolution}

\subsection{Integral scale}

\begin{figure}[h!]\begin{center}
\includegraphics[width=\columnwidth]{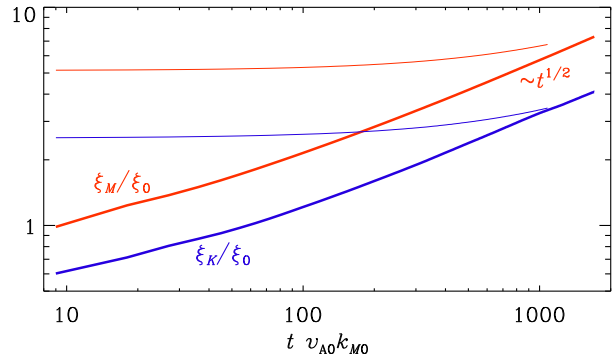}
\end{center}\caption[]{
Kinetic and magnetic integral scales, $\xi_K$ and $\xi_M$, respectively,
for Runs~A (thick lines) and B (thin lines).
}\label{pcomp_kft_QCD_2304_kin}\end{figure}

The decay of MHD turbulence is characterized by the kinetic and
magnetic integral scales.
The kinetic integral scale $\xi_K=k_K^{-1}$ is defined analogously
to the magnetic one $\xi_M=k_M^{-1}$ (given in the Letter), with
\EQ
\kK^{-1}(t)=\int_0^\infty k^{-1} \EK(k,t)\,dk/\EEK(t).
\EN
Both scales grow in time nearly perfectly proportional to each other
like $t^{1/2}$; see \Fig{pcomp_kft_QCD_2304_kin}.
The corresponding decay of kinetic and magnetic energies is proportional
to $t^{-1}$ and is addressed in \Sec{Mach} of this Supplemental Material,
where we plot $\urms=(2\EEK)^{1/2}$ and $\vA=(2\EEM)^{1/2}$.

In Table~1 of Ref.~\cite{SM_Cam14}, Campanelli has summarized the decay laws
for various subinertial range scalings using both helical and non-helical
MHD turbulence.
The scaling exponent for the magnetic integral scale,
$\xiM=\kM^{-1}(t)=\int k^{-1} \EM(k,t)\,dk/\EEM(t)$,
is around 1/2 for most of the different cases.
It emerges quite generically from scaling arguments
first derived by \cite{SM_Ole97}.
Kalelkar \& Pandit \cite{SM_KP04} find $\xiM\sim t^{1/2}$ for
an initial spectrum $\EM\sim k$, but the same scaling also
emerges for other initial power laws, as has been demonstrated by numerous
simulations \cite{SM_kbtr10,SM_Tevzadze:2012kk,SM_Kahniashvili:2012uj}.
The reason for this is that the decay properties depend mainly on nature
of the turbulence (being either hydrodynamic, hydromagnetic without
helicity, or with helicity).

\subsection{Mach \& Alfv\'en numbers}
\label{Mach}

\begin{figure}[h!]\begin{center}
\includegraphics[width=\columnwidth]{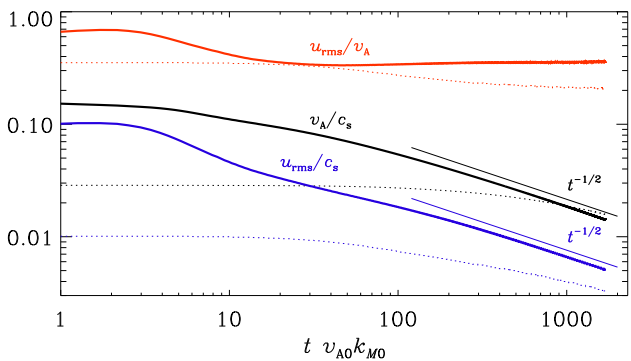}
\end{center}\caption[]{
Mach number $\urms/\cs$ (black),
Alfv\'enic Mach number $\urms/\vA$ (red), and the ratio $\vA/\cs$ (blue)
for Runs~A (thick, solid lines) and B (dotted lines).
}\label{pcomp_ts}\end{figure}

In \Fig{pcomp_ts}, we plot the evolution of the Mach number $\urms/\cs$,
the Alfv\'en number $\urms/\vA$, and the ratio $\vA/\cs$,
where $\urms$ and $\vA$ are the rms values of velocity and magnetic field
(density is approximately constant), and $\cs=\const$ is the isothermal
sound speed.

Both $\urms$ and $\vA$ decay in time proportional to $t^{-1/2}$,
so the kinetic and magnetic energies decay like
$\EEK(t)=\urms^2/2\propto t^{-1}$ and $\EEM (t)=\vA^2/2\propto t^{-1}$.
Earlier work \cite{SM_Kahniashvili:2012uj} resulted in a decay law proportional
to $t^{-0.9}$, but this departure from the $t^{-1}$ law is likely
a consequence of insufficient scale separation; $k_0/k_1$ is now 60
compared to 15 previously.

\subsection{Loitsyansky invariant}
\label{Loitsyansky}

In hydrodynamic turbulence, using the constancy of the Loitsyansky invariant,
\EQ
{\cal L}=\int \rr^2\bra{\uu(\xx)\cdot\uu(\xx+\rr)}\,d\rr \propto \ell^5 u_\ell^2,
\EN
with typical velocity $u_\ell$ on scale $\ell$,
Kolmogorov argued on dimensional grounds that the kinetic energy
should decay like $\EEK\propto{\cal L}^{2/7}t^{-10/7}$.
This decay law is close to the $t^{-5/4}$ scaling found in experiments
\cite{SM_KCM03} and simulations, including those using the
{\sc Pencil Code} \cite{SM_HB04}.

This result has been generalized to MHD by Davidson \cite{SM_Dav00},
although no numerical confirmation of this has been mentioned
in subsequent reviews \cite{SM_Dav10}.
On the other hand, if the decay is governed by viscosity,
dimensional arguments suggest $\EEK\propto\nu t^{-1}$ and
$\ell\propto(\nu t)^{1/2}$, both of which appear consistent
with our simulations.
If that is the case, we should expect ${\cal L}$ to grow with time
like ${\cal L}\propto\nu^{5/2}t^{1/2}$.


\begin{figure*}[t!]
\includegraphics[width=.32\textwidth]{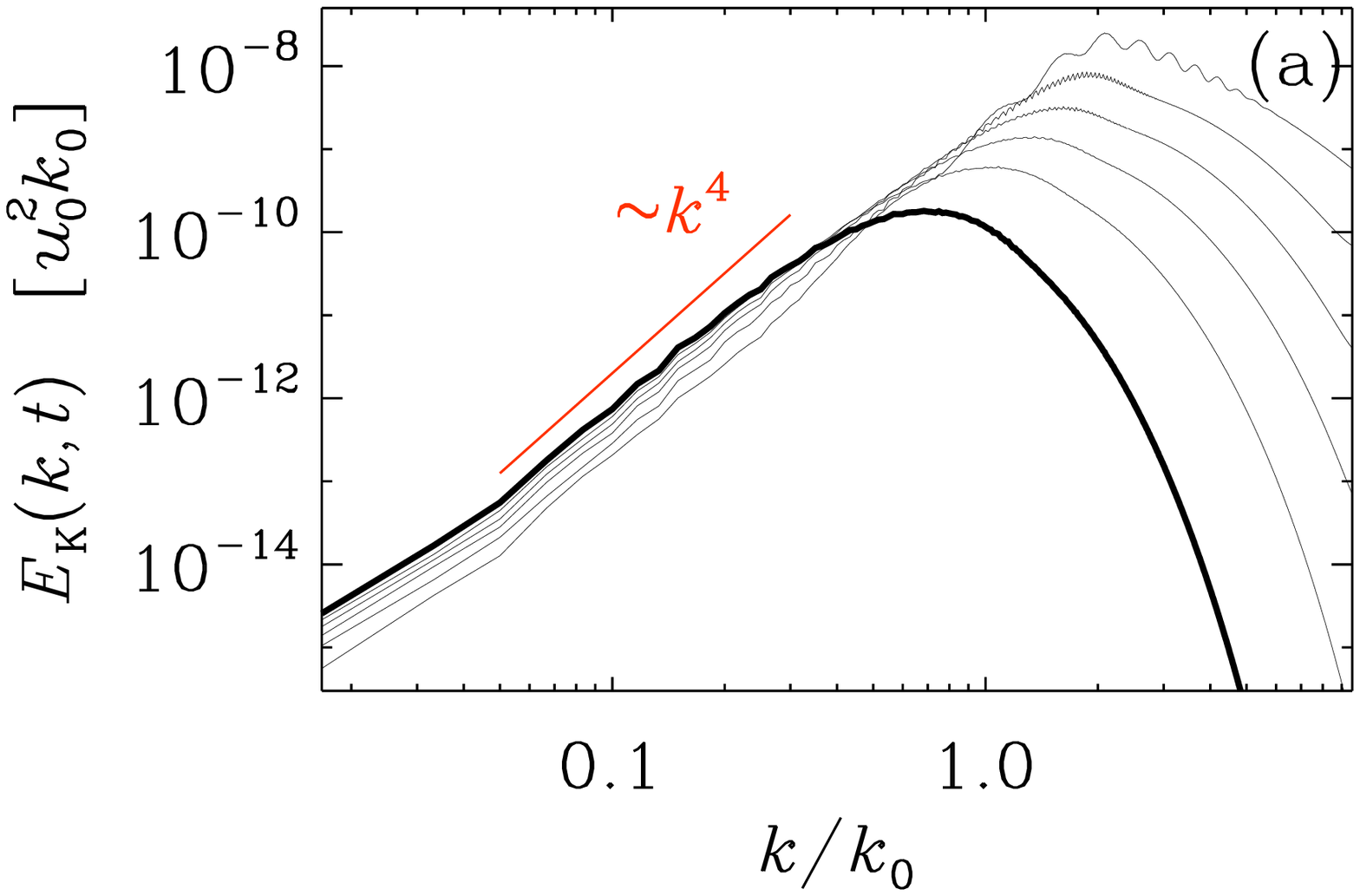}
\includegraphics[width=.32\textwidth]{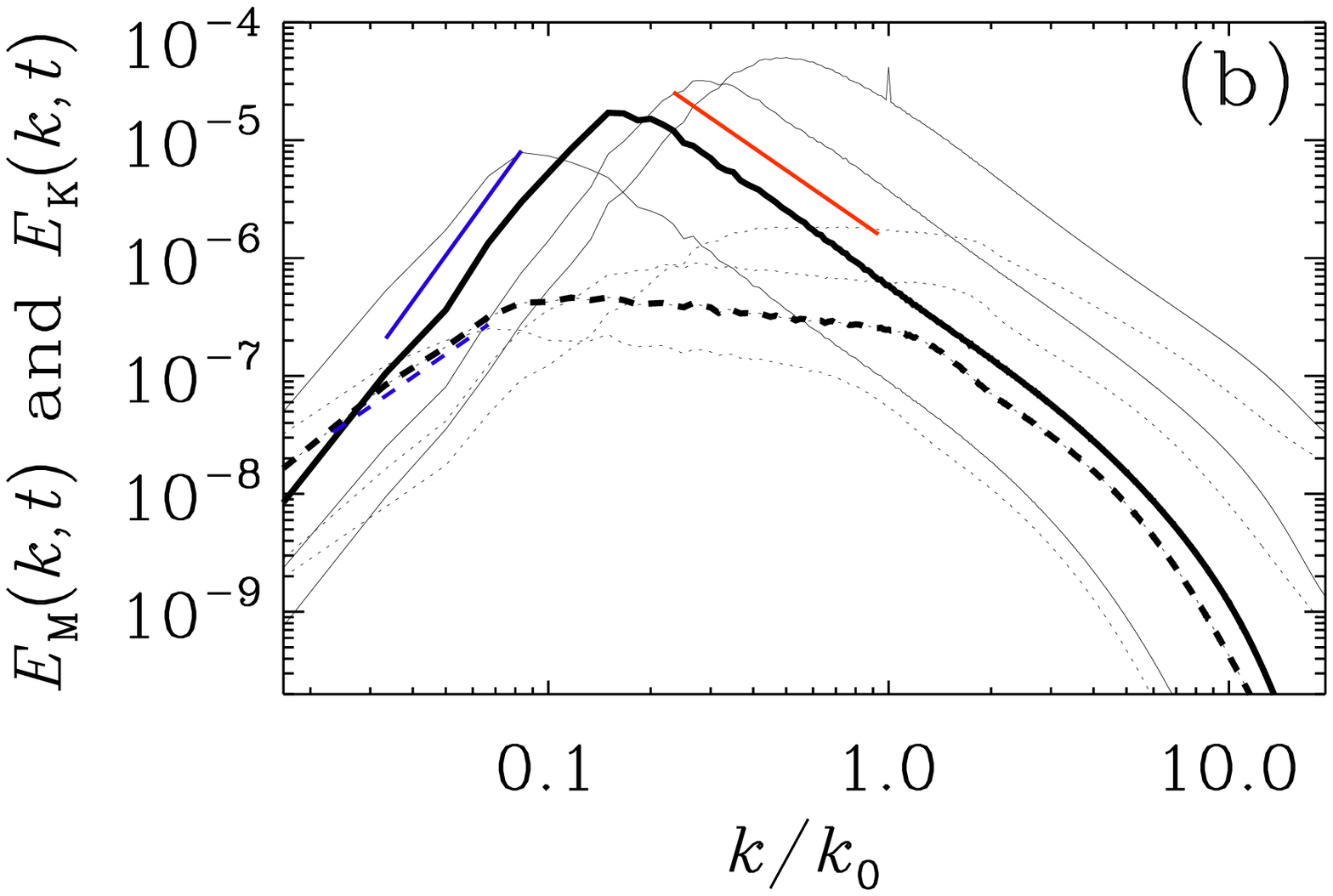}
\includegraphics[width=.32\textwidth]{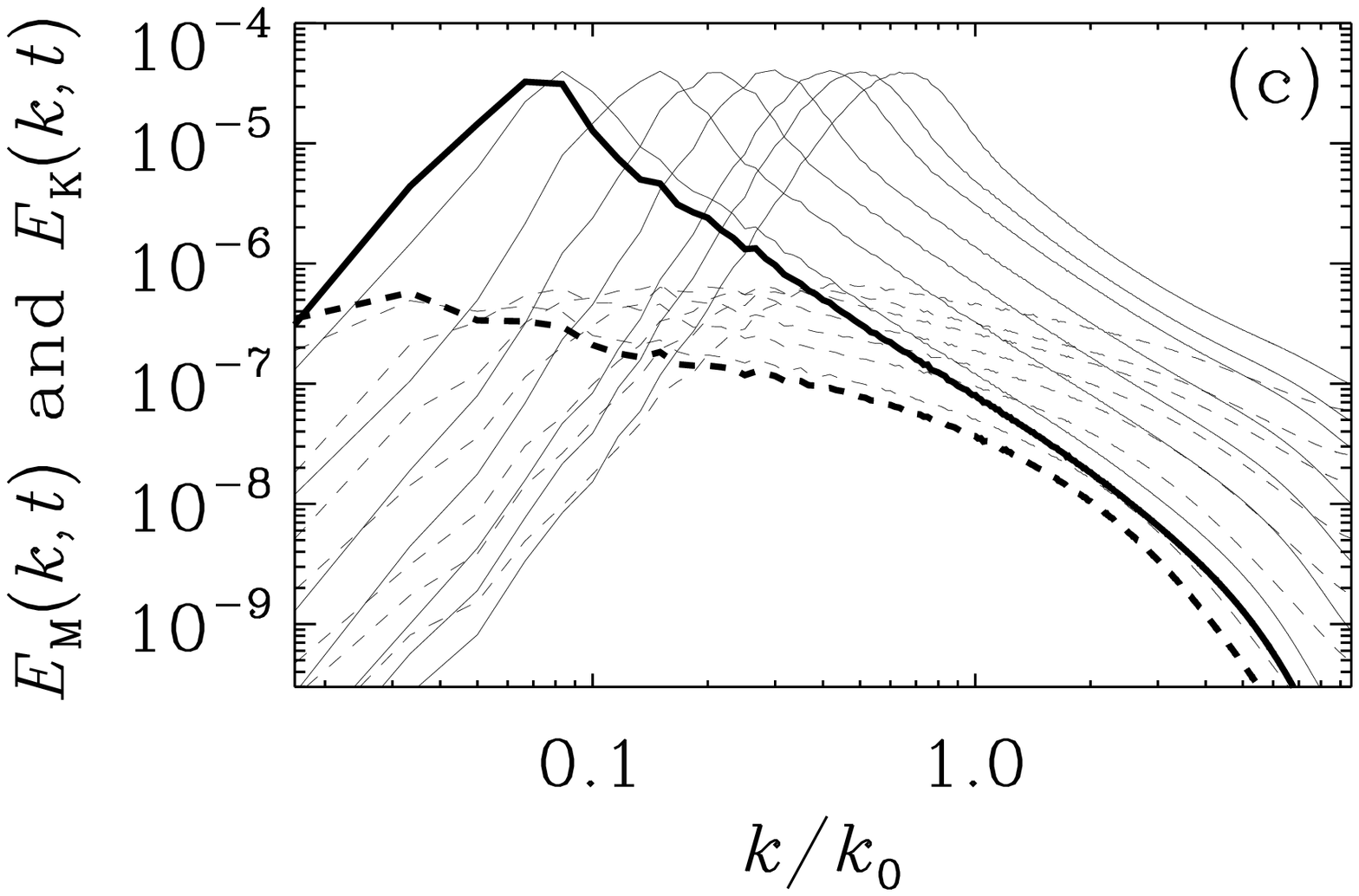}
\\
\includegraphics[width=.32\textwidth]{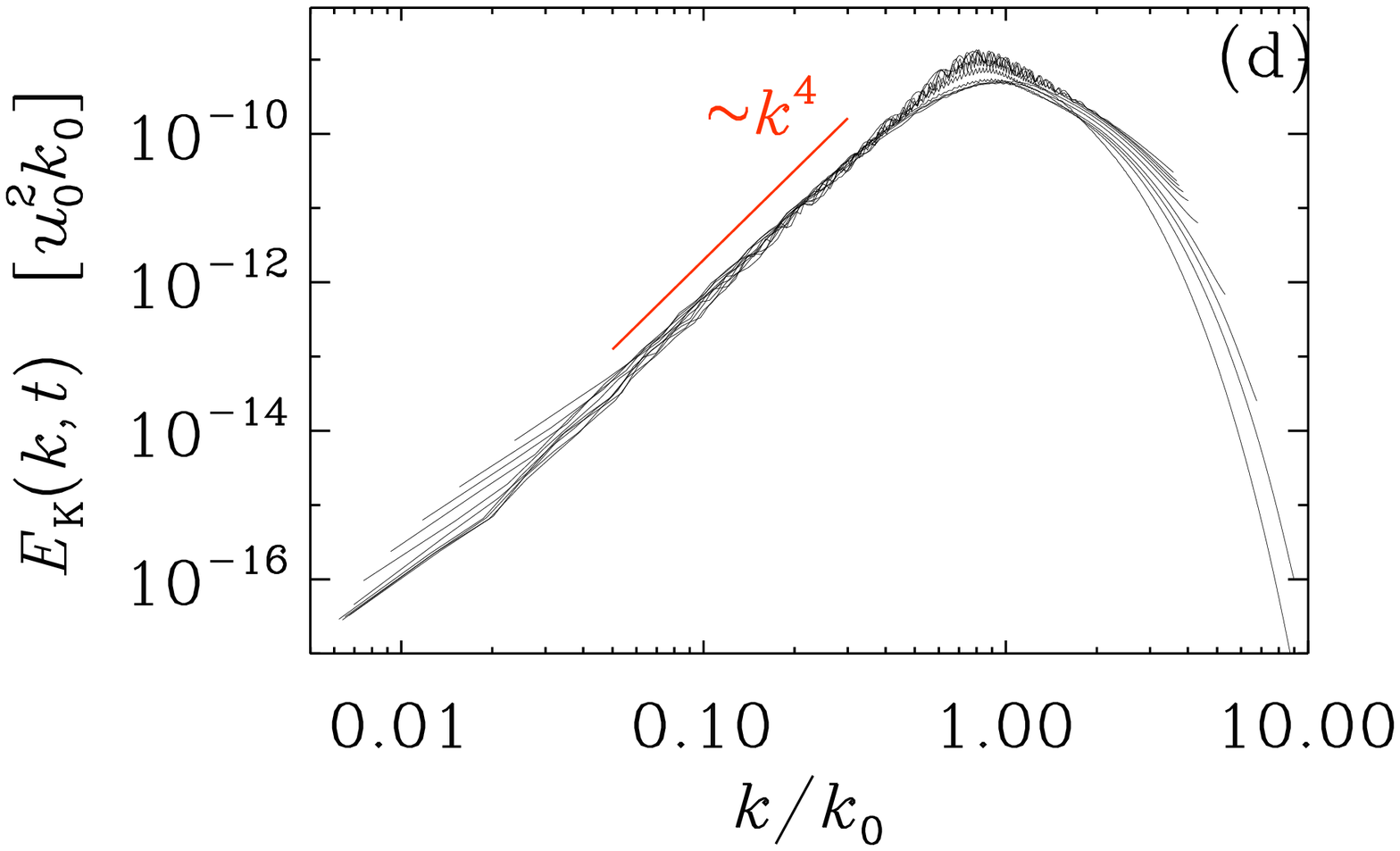}
\includegraphics[width=.32\textwidth]{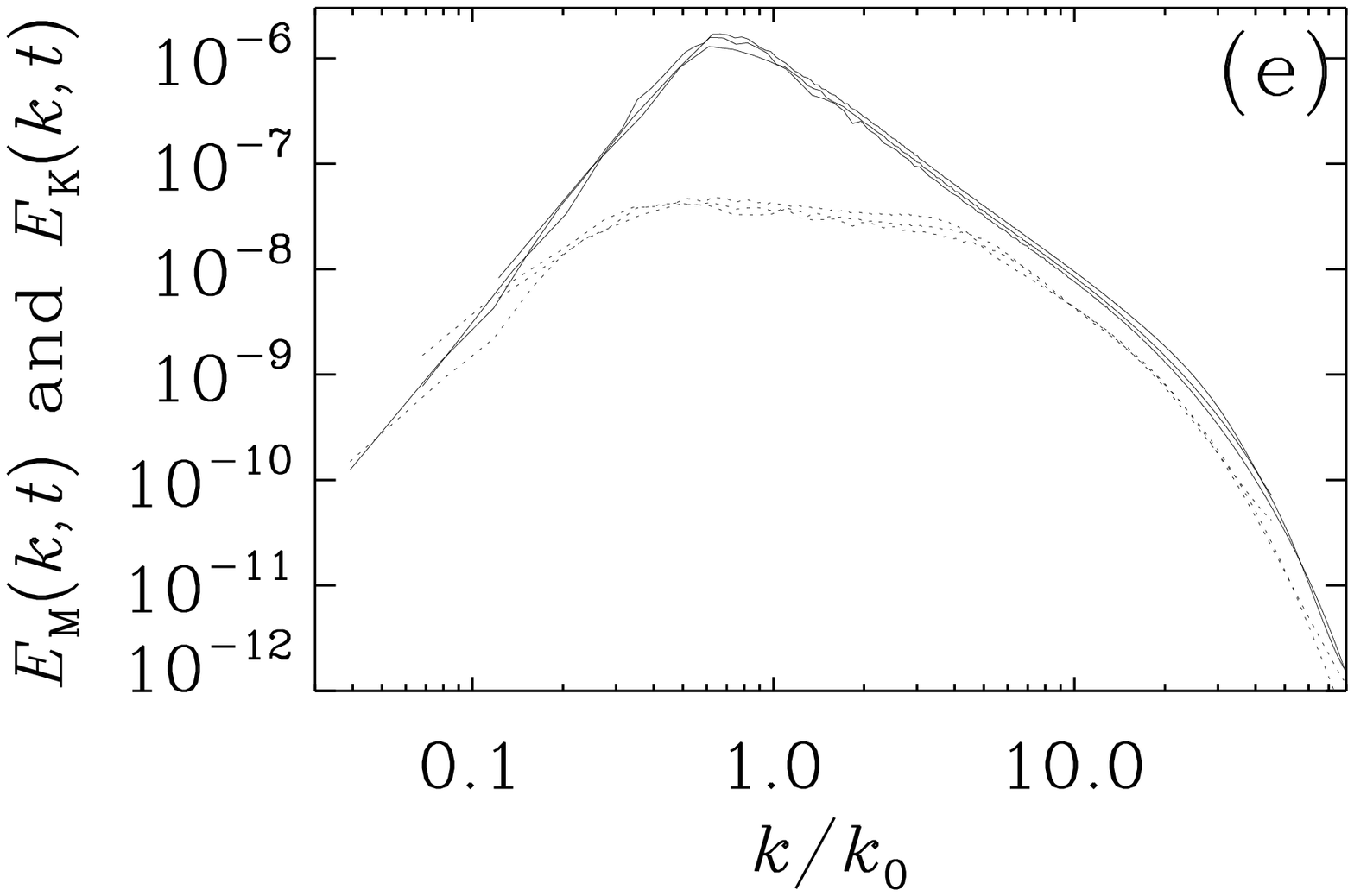}
\includegraphics[width=.32\textwidth]{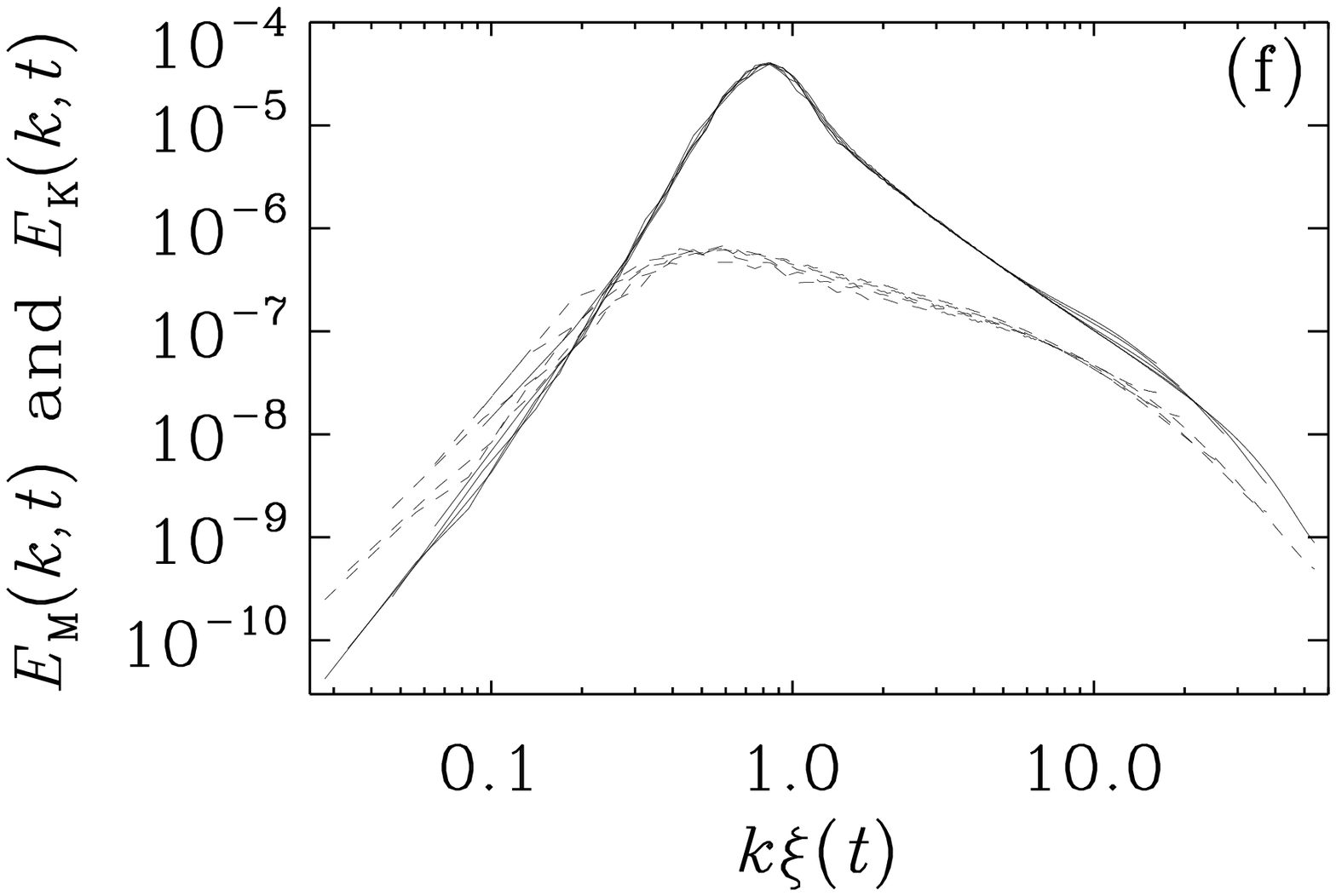}
\\
\caption[]{
Kinetic energy spectra in a hydrodynamic simulation (a),
compared with magnetic (solid) and kinetic (dashed) energy spectra
in a hydromagnetic simulation without helicity (b), and with (c).
Panels (d)--(f) show the corresponding collapsed spectra obtained
by using $\beta=3$ (d), $\beta=1$ (e), and $\beta=0$ (f).
}\label{pkt1152_Kol1152_k4b_120k0}
\end{figure*}

In \Tab{TLoi} we give values of ${\cal L}$ and $\bra{\uu^2}$
for a low resolution run ($64^3$ mesh points) and
a down-sampled one of Run~A of the Letter ($2304^3$ mesh points).

\begin{table}[htb]\caption{
Values of ${\cal L}$ and $\bra{\uu^2}$
for a low resolution run ($64^3$ mesh points) and
a down-sampled one of Run~A of the Letter ($2304^3$ mesh points).
}\vspace{12pt}\centerline{\begin{tabular}{rrcc}
resol.   & $t\;$ & ${\cal L}$ & $\bra{\uu^2}$ \\
\hline
  $64^3$ &  40 & $+2.1\times10^{-7}$ & $8.0\times10^{-3}$ \\
         & 100 & $-5.4\times10^{-7}$ & $7.0\times10^{-5}$ \\
         & 250 & $-8.4\times10^{-7}$ & $1.1\times10^{-5}$ \\
\hline
$2304^3$ &  50 & $-4.4\times10^{-9}$ & $1.0\times10^{-4}$ \\
         & 150 & $-1.1\times10^{-8}$ & $3.3\times10^{-5}$
\label{TLoi}\end{tabular}}
\end{table}

\Tab{TLoi} suggests that $|{\cal L}|$ has indeed a tendency
to increase with time.
This would support our argument above in favor of viscously
dominated decay behavior.
On the other hand, as the resolution is increased by a factor of 36,
${\cal L}$ deceases by about two orders of magnitude while $\urms$
stays about the same.
This would be consistent with ${\cal L}$ converging to zero and
therefore not being able to constrain the decay.

\subsection{Comparison with hydrodynamics}

In the absence of magnetic fields, there is purely hydrodynamic
decay without the mutual interaction between two energy reservoirs.
This leads to a steeper decay law for kinetic energy and a slower
growth of the integral scale, as can be easily be verified by
applying an initial $k^4$ spectrum for the kinetic energy.
This is shown in \Fig{pkt1152_Kol1152_k4b_120k0}, where we compare
the decay spectra for hydrodynamic turbulence with nonhelical and
helical magnetohydrodynamic turbulence.

\subsection{Scaling behavior}

According to the Olesen scaling law \cite{SM_Ole97}, both kinetic and
magnetic energies should decay like
\EQ
\EK(k,t)\sim\EM(k,t)\sim k^\alpha \psi(k^{(3+\alpha)/2}t).
\label{Olesen}
\EN
Integrating over $k$ yields the decay law of the energies as
\EQ
\EEK(t)=\int\EK(k,t)\,d k\sim \int k^\alpha \psi(k^{(3+\alpha)/2}t)\,d k.
\EN
Introducing $\kappa=k t^q$ with $q=2/(3+\alpha)$, we have
\EQ
\EEK(t)\sim t^p \int \kappa^\alpha \psi(\kappa)\,d\kappa,
\EN
where $p=(1+\alpha)q$.
The integral scales like $\kK\sim t^q$ with $q=2/(3+\alpha)$.
Several parameter combinations are given in \Tab{TOle}.

\begin{table}[htb]\caption{
Parameter combinations of
$q=2/(3+\alpha)$, $p=2(1-q)$ and $\alpha=2/q-3$.
Note that $10/7\approx1.43$ and $2/7=0.286$.
}\vspace{12pt}\centerline{\begin{tabular}{crcl}
$\!\!\alpha$ $(=\beta)\!\!$ & $p\;\;$ & $q$ & physics \\
\hline
4  &  $10/7$ & $2/7$ & ${\cal L}=\const$\\
3  &   $8/6$ & $2/6$ \\
2  &   $6/5$ & $2/5$ \\
1  &   $4/4$ & $2/4$ & $\bra{\AAA_{\rm 2D}^2}=\const$ ?\\
0  &   $2/3$ & $2/3$ & $\bra{\AAA\cdot\BB}=\const$ \\
\label{TOle}\end{tabular}}
\end{table}

The numbers in \Tab{TOle} suggest that different subinertial range scalings
$\propto k^\alpha$ correspond to different exponents $p$ and $q$.
Instead, we now argue that the subinertial range scaling is not
characterized just by $\alpha$, but mainly by the slope of
$\psi(k)$ for small values of $k$.
To demonstrate this, let us begin by noting that, in practice,
it is more accurate to relate the change in the amplitude
to the change in the correlation length instead of the time,
because it always takes a while to establish asymptotic scaling behavior.
Therefore, instead of using \Eq{Olesen}, we prefer the following
alternative formulation by extracting just a time-dependent factor
from the spectrum an write \cite{SM_CHB01}
\EQ
\EM(k,t)\sim \xi^\beta \phi(k\xi),
\EN
with $\xi(t)\sim t^q$ standing either for $\xiM$ (in the magnetic case)
or for $\xiK$ (in the hydrodynamic case).
The exponent $q$ can be determined by dimensional arguments,
e.g.\ by assuming ${\cal L} = \const = U^2 L^5$, which implies
\EQ
{\cal L} \sim u^2 l^5 \sim l^7\tau^{-2} \to q=2/7,\; p=10/7 \; (\beta=4).
\label{calL}
\EN
Alternatively, in the presence of helical magnetic fields,
$\bra{\AAA\cdot\BB}$ is conserved.
Again, from dimensional arguments we find
\EQ
\bra{\AAA\cdot\BB} \sim u^2 l \sim l^3\tau^{-2} \to q=2/3,\; p=2/3 \; (\beta=0).
\EN
Finally, in nonhelical hydromagnetic turbulence we have $q=1/2$, which
suggests that a quantity with the dimensions $u^2 l^2$ should be constant.
Since $\bra{\AAA^2}/\mu_0\rho_0$ has such dimensions $u^2 l^2$,
we must carefully reassess our previous findings suggesting that the flow
is fully three-dimensional.
Nevertheless, tentatively one can state
\EQ
\bra{\AAA_{\rm 2D}^2} \sim u^2 l^2 \sim l^4\tau^{-2} \to q=1/2,\; p=1 \; (\beta=1),
\label{AA2}
\EN
where $\AAA_{\rm 2D}$ is the gauge that aligns $\AAA$ with the intermediate
rate-of-strain vector; see \Sec{SquaredPotential}.
\Eqss{calL}{AA2} a decay law $\EM\sim t^{-p}$ with $p=(1+\beta)\,q$,
where $\beta$ is a parameter that is usually {\em not} associated with
the subinertial range scaling exponent $\alpha$.
Formally, however, we find that $\alpha = \beta$, so the parameter
combinations in \Tab{TOle} still apply with $\alpha = \beta$, but this new
formulation does not imply anything about the subinertial range scaling.

\section{Inverse transfer}

The growth of spectral energy at small wavenumbers is
cautiously referred to as inverse {\em transfer}.
By contrast, in an inverse {\em cascade} there is a $k$-independent
flux of some quantity (e.g., of magnetic helicity in three-dimensional
hydromagnetic turbulence) toward progressively smaller $k$.
An inverse cascade is known to exist in three dimensional
hydromagnetic turbulence if there is helicity.
Also two-dimensional hydrodynamic turbulence is known
to exhibit inverse cascade behavior.

While any of these mechanisms could in principle play a
role in explaining the behavior seen in the Letter \cite{SM_BKT14},
there may also be completely different mechanisms that could
potentially explain the growth of spectral magnetic energy
at large length scales.
In order to narrow down possible reasons for the inverse
transfer found in our simulations, we begin by discussing
the concepts of eddy noise and the unwinding of magnetic fields.
We also determine the magnetic helicity as well as various
other helicities and examine whether the flow could be locally
two-dimensional, which might provide yet another cause of
inverse transfer, in which case one could appeal to the inverse cascade
of $\bra{\AAA_{\rm 2D}^2}$, where $\AAA_{\rm 2D}$ is the magnetic vector
potential perpendicular in a gauge in which it is locally aligned
with the intermediate eigenvector of the rate-of-strain tensor.

\subsection{Eddy noise}

Eddy noise has been mentioned as a mechanism that brings energy from
the unresolved scales into resolved scales \cite{SM_Bae08,SM_Sen12}.
The physics of this mechanism is obscured by the fact that it
referred originally to numerical artifacts.
In particular, it is not obvious that it really leads to a spectral
increase of power rather than just a preferential decay at small scales.
Of course, at a descriptive level, the concept of eddy noise may
be the similar to what we find, although there was never a clear
demonstration of the resulting spectral evolution.

\subsection{Unwinding magnetic fields}

One could imagine that the unwinding of a magnetic field
leads to the conditions seen in the Letter.
Again, this is only a descriptive concept, but one could think
of addressing this question through a numerical experiment.
Winding up an externally imposed magnetic field leads to
magnetic flux expulsion, as first shown by Weiss \cite{SM_Wei66}.
We now ask about the dynamical behavior of such a magnetic
field after the diving force is turned off.

\begin{figure}[h!]\begin{center}
\includegraphics[width=.49\columnwidth]{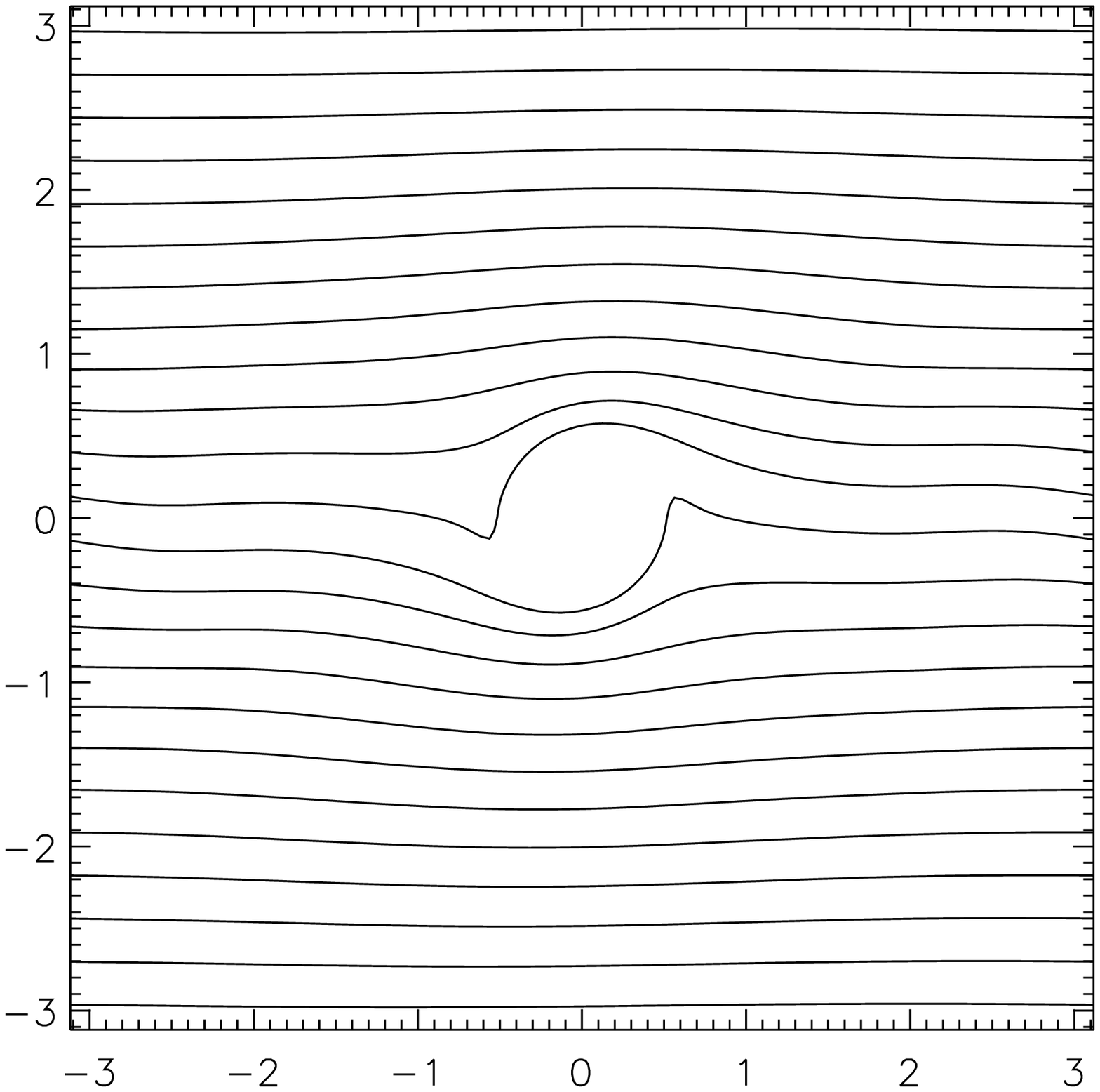}
\includegraphics[width=.49\columnwidth]{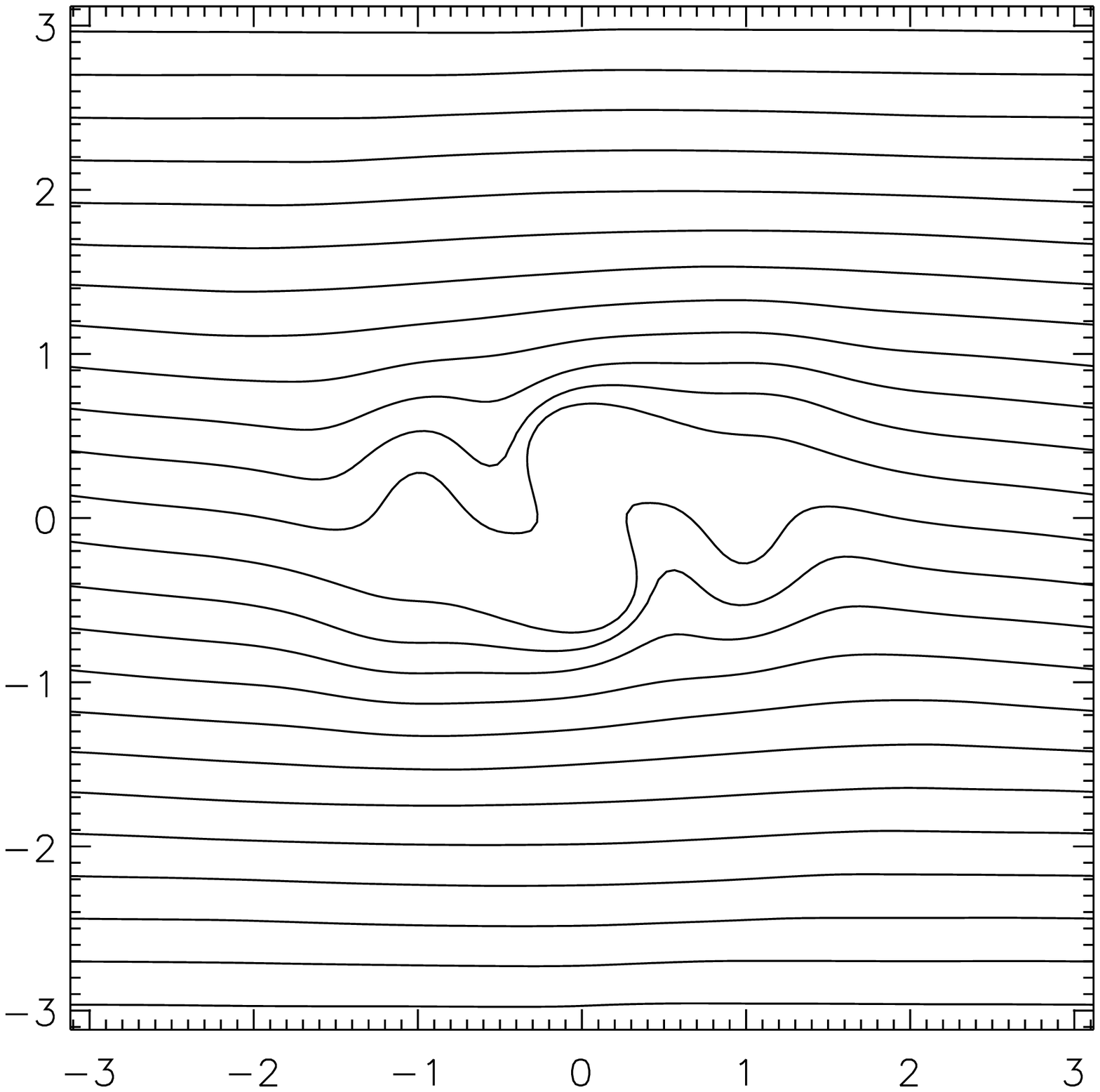}
\end{center}\caption[]{
Field lines in a two-dimensional simulation in the presence
of a forced hydromagnetic eddy at an intermediate time (left)
and a later time (right) when the eddy has been altered
by the Lorentz force in a time-dependent manner.
}\label{pvar_128_t100}\end{figure}

\begin{figure}[h!]\begin{center}
\includegraphics[width=\columnwidth]{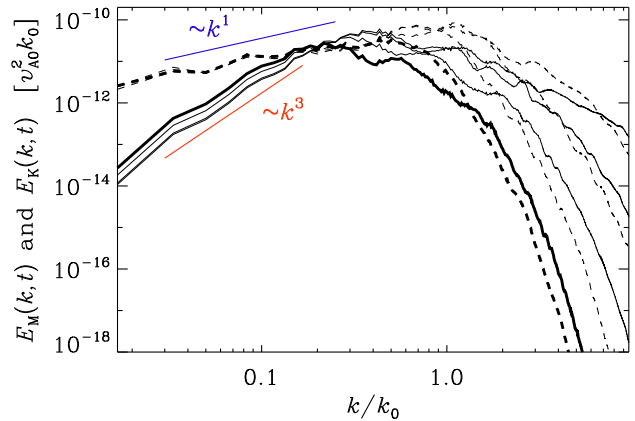}
\end{center}\caption[]{
Kinetic and magnetic energy spectra.
The two red lines are proportional to $k^{1}$ and $t^{3}$, respectively.
}\label{pkt2304_2304b_Bvortex_off}\end{figure}

There are several complications with this seemingly simple idea.
First, the original problem of Weiss was not dynamic, but kinematic.
Allowing the flow to be a self-consistent solution of the momentum
equation leads to more complicated behavior, as demonstrated
in \Fig{pvar_128_t100}, where the driving force is given by
$\ff=\nab\times\psi\zzz$ with $\psi=\cos k_1 x\cos k_1 y\exp(-r^2/2R^2)$,
and $r^2=x^2+y^2$.
Second, turning off the driving leads to propagating Alfv\'en
waves and long-term oscillations at large scales,
which is very different from what is seen in the Letter.
However, this behavior can easily be removed by also turning off
the imposed magnetic field.
Finally, the eddy shown in \Fig{pvar_128_t100} is too big
($R=0.1$) and there would be no extended subinertial range.
Thus, we now repeat this experiment with a much smaller eddy of radius
$R=0.002$.
The resulting spectral evolution is shown in \Fig{pkt2304_2304b_Bvortex_off}.

In this experiment, the initial magnetic field is a global one
and it therefore unconstrained by causality.
It turns out that the magnetic energy now has a $k^3$ subrange
at large scales.
Interestingly, as the magnetic field decays, there is
a slight decay also at large length scales, so there
is no inverse transfer in such a setup.

\subsection{Helicities for Runs~A and B}

In \Tab{Thel} we list various helicities for Runs~A and B.
The normalized kinetic helicity
$\bra{\oo\cdot\uu}/\omega_{\rm rms}u_{\rm rms}$
and the normalized cross helicity
$2\bra{\uu\cdot\bb}/(u_{\rm rms}^2+b_{\rm rms}^2)$
are less than 0.4\%,
the normalized current helicity
$\bra{\jj\cdot\bb}/j_{\rm rms}b_{\rm rms}$ is less than 2\%,
and the normalized magnetic helicity
$k_1\bra{\aaaa\cdot\bb}/b_{\rm rms}^2$ is less than 1\%.
In this Letter, it was shown that the level of magnetic helicity
was small enough so as not to constrain (i.e., enhance) the growth
of the integral scale.
Also the other helicities appear to be small enough for being important
in explaining the inverse transfer.

\begin{table}[htb]\caption{
Various helicities for Run~A of the Letter.
}\vspace{12pt}\centerline{\begin{tabular}{llr}
quantity & expression & value \\
\hline
kinetic hel. & $\bra{\oo\cdot\uu}/\omega_{\rm rms}u_{\rm rms}$ & $0.00364$ \\
current hel. & $\bra{\jj\cdot\bb}/j_{\rm rms}b_{\rm rms}$      & $0.01693$ \\
cross hel. & $2\bra{\uu\cdot\bb}/(u_{\rm rms}^2+b_{\rm rms}^2)$ & $-0.00318$ \\
magnetic hel. & $k_1\bra{\aaaa\cdot\bb}/b_{\rm rms}^2$ & $ 0.00976$ \\
\label{Thel}\end{tabular}}\end{table}

\subsection{Projections onto strain tensor}

To examine whether there is a tendency for the turbulence to become
locally two-dimensional, we compute the rate-of-strain tensor,
\EQ
s_{ij}=\half\left(u_{i,j}+u_{j,i}\right).
\EN
Since $s_{ij}$ is symmetric, it has three real eigenvalues,
$\lambda_i$ for $i=1$, 2, and 3.
They are traditionally ordered such that
\EQ
\lambda_1 < \lambda_2 < \lambda_3.
\EN
The corresponding eigenvectors are called $\eee_i$.

If the flow was incompressible, their sum would vanish, which
is here also approximately the case.
The largest eigenvalue $\lambda_3$ corresponds to stretching
in the direction $\eee_3$, and the most negative one, $\lambda_1$,
corresponds to compression in the direction $\eee_1$.

\begin{figure}[h!]\begin{center}
\includegraphics[width=\columnwidth]{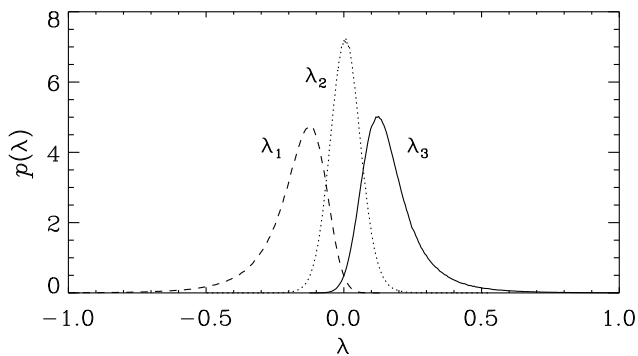}
\end{center}\caption[]{
PDF of the eigenvalues of the rate-of-strain tensor.
Note that the intermediate ones are not vanishing,
as expected for two-dimensional turbulence.
}\label{peigenvals}\end{figure}

It has been known for some time \cite{SM_Ker85,SM_Ash87} that in isotropic
turbulence the vorticity vector tends to be aligned with the direction
$\eee_2$ and is therefore normal to the plane where the flow would be
two-dimensional.
If the turbulence was perfectly two-dimensional, the intermediate
eigenvalue of the rate-of-strain tensor would vanish.
This is however not the case; see \Fig{peigenvals}, where we plot
probability density functions (PDFs) of the three eigenvalues.

\begin{figure}[h!]\begin{center}
\includegraphics[width=\columnwidth]{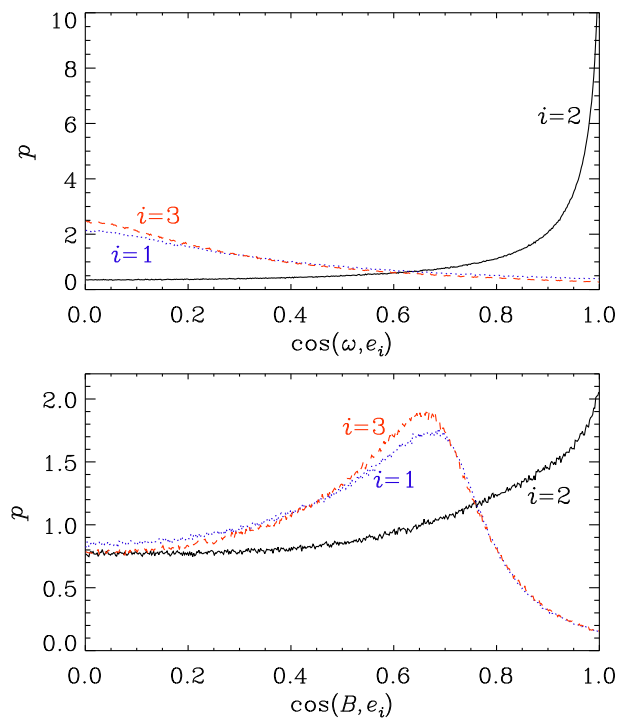}
\end{center}\caption[]{
Alignment of $\oo$ and $\BB$ with the eigenvectors of the
rate-of-strain tensor.
}\label{palign}\end{figure}

In the present simulations, we see all the usual characteristics
of three-dimensional MHD turbulence where the vorticity vector $\oo$
is aligned with the eigenvector $\eee_2$.
Also the magnetic field $\BB$ is aligned with $\eee_2$; see \Fig{palign}.
Here, the PDFs $p(\cos\phi)$ are normalized such that
\EQ
\int_0^1 p(\cos\phi)\,d\cos\phi=1.
\EN
Furthermore, while $\oo$ is perpendicular to $\eee_1$ and $\eee_3$,
the angle between $\BB$ and both $\eee_1$ and $\eee_3$ is about $45^\circ$
(see lower panel of \Fig{palign}), which was first found in MHD shear flows;
see Ref.~\cite{SM_BNST95}, who interpreted their finding as alignment
with the direction of the overall shear.
In this connection we recall that a shear flow can be decomposed into
a rotational and a straining motion \cite{SM_B95}.
The rotational motion is not captured by the strain tensor.
The directions of compression and stretching are then at $45^\circ$
angles with respect to the direction of the shear \cite{SM_BJNRST96}.
Similar results have recently also been obtained in Ref.~\cite{SM_SPS14}.

Next, we compute the projections of various vectors $\ssigma$ onto $s_{ij}$:
\begin{equation}
{\sf s}_\sigma=\bra{\sigma_i s_{ij} \sigma_j}/\bra{\ssigma^2},
\end{equation}
where $\ssigma$ is either $\uu$, $\oo$, $\BB$, or $\JJ$.
For $\oo$ and $\BB$, these values quantify the production of
$\oo$ and $\BB$, respectively, but for the other quantities
no such interpretation exists.
In \Tab{Tsigma} we give the mean and rms values of the ${\sf s}_\sigma$.
Here we also compare with the projection of $\AAA$ onto the direction
$\eee_2$, i.e., $\AAA\to\eee_2(\AAA\cdot\eee_2)$, and therefore
${\sf s}_A=\bra{\lambda_2(\AAA\cdot\eee_2)^2}/\bra{(\AAA\cdot\eee_2)^2}$.

\begin{table}[htb]\caption{
Mean and rms values of the normalized projections ${\sf s}_\sigma$ with
$\ssigma=\uu$, $\oo$, $\BB$, and $\JJ$ onto $s_{ij}$ and a comparison
with the value of ${\sf s}_A$ defined in the text.
}\vspace{12pt}\centerline{\begin{tabular}{lccrcccc}
$\ssigma$ & $\uu$ & $\oo$ & $\BB\;\;$ & $\JJ$ & $\AAA$ \\   
\hline
mean & $ 0.009$ & $ 0.019$ & $-0.001$ & $ 0.005$ & $0.007$ \\
rms  & $ 0.145$ & $ 0.204$ & $ 0.069$ & $ 0.064$ & $0.106$%
\label{Tsigma}\end{tabular}}\end{table}

Note that ${\sf s}_A$ is not particularly small, as one would have
expected for a locally nearly two-dimensional flow and is comparable
to all the other terms.
The average of $\bra{B_i s_{ij} B_j}/\bra{\BB^2}$ is actually the
smallest one among them all.

\subsection{Conservation of squared potential?}
\label{SquaredPotential}

In the Letter, we have argued that the inverse transfer can be
a consequence of the different subinertial range scalings for
kinetic and magnetic energy spectra.
However, as a very different possibility we also have to consider
an inverse cascade due to the approximate invariance of $\bra{\AAA^2}$
in two dimensions.
This would seem surprising given the similar widths of the PDFs of the
intermediate eigenvalue $\lambda_2$ and those of $\lambda_1$ and $\lambda_3$,
which suggests that the flow cannot be regarded as locally two-dimensional.

The quantity $\bra{\AAA^2}$ is obviously gauge-dependent.
However, the relevant gauge is one that aligns $\AAA$ locally with $\eee_2$.
Thus, we use the new vector potential
\EQ
\AAA_{\rm 2D}=\AAA+\nab\Lambda,
\EN
such that
\EQ
0=\eee_2\times\AAA+\eee_2\times\nab\Lambda,
\EN
Taking the curl yields
\EQ
0=\eee_2\cdot\nab\times(\eee_2\times\AAA)
+\eee_2\cdot\nab\times(\eee_2\times\nab\Lambda),
\EN
which can be solved in Fourier space (indicated by hats) to give
\EQA
\nab\Lambda=-\int{\kk\cdot\hatAA-(\eee_2\cdot\kk)(\eee_2\cdot\hatAA)
\over \kk^2-(\eee_2\cdot\kk)^2}\,\kk\, e^{i\kk\cdot\xx}\,d^3k,
\ENA

\begin{table}[htb]\caption{
Values of $\bra{\AAA^2}$, $\bra{\AAA_{\rm 2D}^2}$, and $\bra{\BB^2}$
for a low resolution run ($144^3$ mesh points) and the down-sampled Run~A
of the Letter ($2304^3$ mesh points).
}\vspace{12pt}\centerline{\begin{tabular}{rrccc}
resol.   & $t\;$ & $\bra{\AAA^2}$ & $\bra{\AAA_{\rm 2D}^2}$ & $\bra{\BB^2}$ \\
\hline
 $144^3$ &  10 & $1.8\times10^{-5}$ & $9.7\times10^{-6}$ & $1.8\times10^{-4}$ \\
         &  50 & $1.2\times10^{-5}$ & $6.0\times10^{-6}$ & $8.1\times10^{-5}$ \\
         & 100 & $9.4\times10^{-6}$ & $5.1\times10^{-6}$ & $4.4\times10^{-5}$ \\
         & 200 & $6.6\times10^{-6}$ & $5.0\times10^{-6}$ & $2.0\times10^{-5}$ \\
\hline
$2304^3$ &  50 & $5.3\times10^{-5}$ & $2.0\times10^{-5}$ & $8.2\times10^{-4}$ \\
         & 100 & $4.4\times10^{-5}$ & $9.9\times10^{-6}$ & $4.0\times10^{-4}$ \\
         & 150 & $4.0\times10^{-5}$ & $8.7\times10^{-6}$ & $2.6\times10^{-4}$ \\
\label{TA2}\end{tabular}}
\end{table}

In \Tab{TA2} we summarize the values of $\bra{\AAA^2}$ (in the gauge
used in the code, i.e., the Weyl gauge) and $\bra{\AAA_{\rm 2D}^2}$,
and compare their temporal changes with that of $\bra{\BB^2}$ for a
low resolution run and the down-sampled Run~A of the Letter.
Note first of all that $\bra{\AAA_{\rm 2D}^2}$ is always smaller than
$\bra{\AAA^2}$.
This is expected, because $\AAA$ contains redundant contributions.
Second, $\bra{\AAA_{\rm 2D}^2}$ decays more slowly than $\bra{\BB^2}$,
demonstrating that $\bra{\AAA_{\rm 2D}^2}$ is approximately conserved.
Furthermore, the ratio $(\bra{\AAA_{\rm 2D}^2}/\bra{\BB^2})^{1/2}$
is a length scale of around 0.2 for Run~A, which is well above the
scale Taylor micro scale shown in Fig.~4 of the Letter.
Thus, the typical values of $\bra{\AAA_{\rm 2D}^2}$ may well be significant
for explaining the $\xiM\sim t^{1/2}$ scaling and its approximate
conservation could be responsible for the inverse cascade \cite{SM_Pou78}.

\subsection{Reynolds and Lundquist numbers}

\begin{figure}[h!]\begin{center}
\includegraphics[width=\columnwidth]{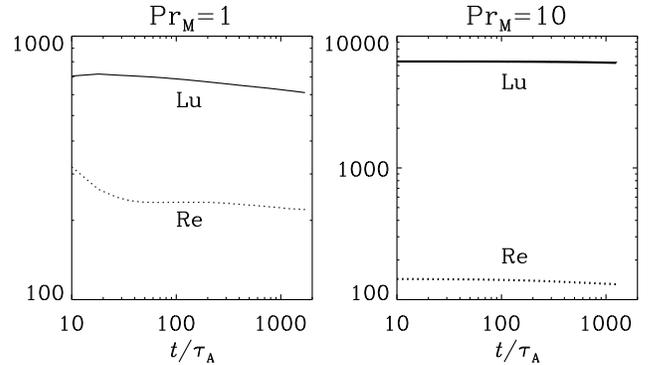}
\end{center}\caption[]{
Instantaneous magnetic and kinetic Reynolds numbers
for Runs~A and B with $\Pm=1$ and 10, respectively.
}\label{pcomp_kft_QCD_2304_tscale}\end{figure}

Since $\urms$, $\vA$, and $\kM$ are all proportional to $t^{-1/2}$
the decay is self-similar in such a way that the Reynolds and Lundquist
numbers, $\Rey=\urms/\nu\kM$ and $\Lu=\vA/\eta\kM$, remain constant.
This is clearly demonstrated in \Fig{pcomp_kft_QCD_2304_tscale},
where we plot $\Rey$ (dotted) and $\Lu$ (solid) both for
Runs~A and B with $\Pm=1$ and 10, respectively.

During the decay, the approximate values of $\Rey$ and $\Lu$ are
230 and 700 for Run~A and 130 and 6300 for Run~B.
As discussed in the Letter, for Run~B the value of $\Lu$ is so
huge that, even though we also have  a large number of mesh points,
one must be concerned about the numerical accuracy of the simulation.
One should note, however, that, because $\Pm=10$ is larger than unity,
most of the energy is dissipated viscously.
Therefore, although $\Lu$ is huge, less magnetic energy needs to be
dissipated than for $\Pm=1$.
This was demonstrated first in Ref.~\cite{SM_B09} for the opposite case
with $\Pm\ll1$ (where Joule dissipation dominates) and later, in
Refs.~\cite{SM_B11,SM_B14}, for $\Pm\gg1$, which is the case relevant here.

\section{Energy transfers}

In the Letter we considered the spectral transfer function
$T_{kpq}=\bra{\JJ^k\cdot(\uu^p\times\BB^q)}$, which governs the
gain of magnetic energy and correspondingly the loss of kinetic energy.
There is also a kinetic energy transfer function $S_{kpq}$,
which describes the transfer between different scales.
It vanishes for $k=p$ and is given by
$S_{kpq}=\rho_0\bra{\uu^k\cdot(\uu^p\times\oo^q)}$,
where $\rho_0=\bra{\rho}$ is the average density;
compressibility effects have been ignored here.
The transfer function $S_{kpq}$ enters only in the kinetic energy equation.
Thus, the magnetic and kinetic energy equations are given by
\begin{equation}
{d\over dt}\bra{\half\BB_k^2}=T_{kpq}-\eta k^2\bra{\BB_k^2},
\label{Eqn1}
\end{equation}
\begin{equation}
{d\over dt}\bra{\half\rho_0\uu_p^2}=-T_{kpq}-S_{kpq}
-\nu p^2\bra{\rho_0\uu_p^2}.
\label{Eqn2}
\end{equation}
Swapping indices $k$ and $p$ in \Eq{Eqn2} yields
\begin{equation}
{d\over dt}\bra{\half\rho_0\uu_k^2}=-T_{pkq}-S_{pkq}
-\nu k^2\bra{\rho_0\uu_k^2}.
\label{Eqn2b}
\end{equation}
To get the total energy at wavenumber $k$, we now add \Eqs{Eqn1}{Eqn2b},
i.e.,
\begin{eqnarray}
\half{d\over dt}\bra{\BB_k^2+\rho_0\uu_k^2}&\!=\!&T_{kpq}-T_{pkq}-S_{pkq}
\nonumber \\
&\!-\!&\eta k^2\bra{\BB_k^2}-\nu k^2\bra{\rho_0\uu_k^2}.
\end{eqnarray}
The total energy has contributions from all $p$ and $q$ and it is of interest
to separate between those that are larger and smaller than $k$,
so we write
\begin{eqnarray}
\half{d\over dt}\bra{\BB_k^2+\rho_0\uu_k^2}
&\!=\!&\Pi_{p\geq k}^{q\geq k}+\Pi_{p\geq k}^{q < k}
+\Pi_{p < k}^{q\geq k}+\Pi_{p < k}^{q < k}
\nonumber \\
&\!-\!&\eta k^2\bra{\BB_k^2}-\nu k^2\bra{\rho_0\uu_k^2}.
\end{eqnarray}
where
\newcommand{\gl}{\stackrel{<}{>}}
\begin{equation}
\Pi_{p \gl k}^{q \gl k}
=\sum_{p\gl k}\sum_{q\gl k}\left(T_{kpq}-T_{pkq}-S_{pkq}\right).
\label{TransferFunct}
\end{equation}

\begin{figure}[h!]\begin{center}
\includegraphics[width=\columnwidth]{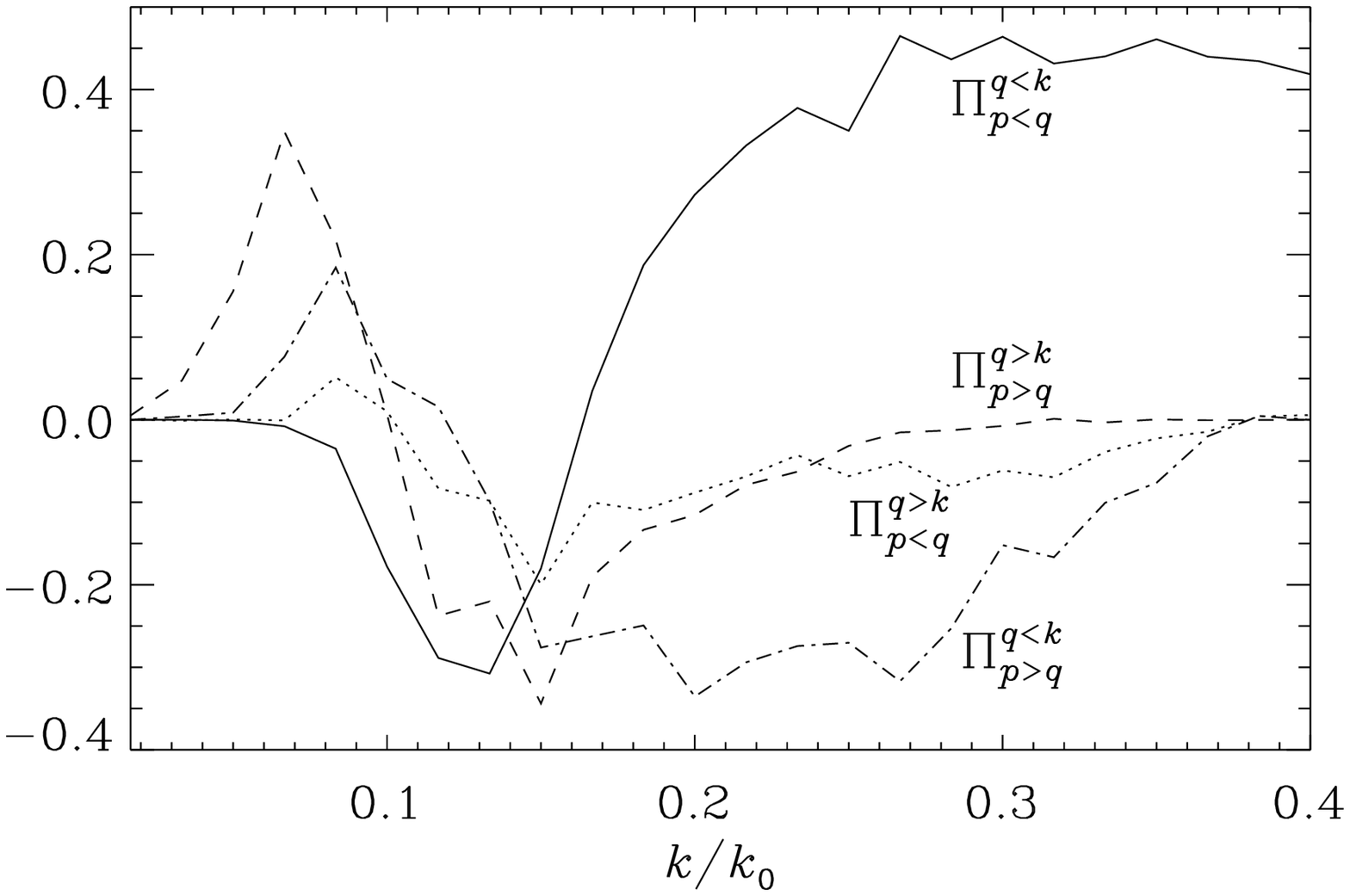}
\end{center}\caption[]{
Energy transfer functions defined in \Eq{TransferFunct}.
}\label{pST_kpq_transfer}\end{figure}

The results in \Fig{pST_kpq_transfer} show that $\Pi_{p < k}^{q < k}$ is
positive for $k/k_0 > 0.25$, demonstrating that there is a gain of total energy
at wavenumbers $k/k_0 > 0.25$ from interactions with smaller wavenumbers.
This shows that there is forward transfer at those wavenumbers.
Furthermore, for $k/k_0 > 0.25$, the spectral transfer is approximately
independent of $k$, as expected for a proper forward cascade.
There is also a short range $0.15 < k/k_0 < 0.3$, where
$\Pi_{p > k}^{q < k}$ is negative.
This suggests the existence of inverse transfer resulting from
mixed interactions of $p>k$ and $q<k$.
In our simulations, $|T_{kpq}|$ dominates over $|S_{kpq}|$, so the
dominant contribution to wavenumbers $q$ comes from the magnetic field.

\section{Concluding remarks}

The decay of magnetically dominated MHD turbulence is a rich field,
sharing several similarities with the case in which the magnetic field
is dynamo-generated, for example the alignment properties with the
eigenvectors of the rate-of-strain tensor.
In the Letter, we have focussed on the inverse transfer properties
that were previously only known for helical MHD turbulence.
This is a new and exciting result that has now been confirmed by
two additional independent groups \cite{SM_BL14,SM_Zra14}.
The results presented in the Supplemental Material support the
robustness of this result and suggest that the inverse transfer is not
explained by other previously studied mechanisms.


\vfill\bigskip\noindent\tiny\begin{verbatim}
$Header: /var/cvs/brandenb/tex/tina/Decay/both.tex,v 1.4 2014/11/28 19:12:22 brandenb Exp $
\end{verbatim}

\end{document}